\documentclass[sigconf,natbib=false,nonacm]{acmart}

\renewcommand\footnotetextcopyrightpermission[1]{} % removes footnote with conference information in first column
\usepackage{algorithm}
\usepackage{algpseudocode}
\usepackage{graphicx}
\usepackage{subcaption}
\usepackage{textcomp}
\usepackage{cuted}
\usepackage{xcolor}
\usepackage{glossaries}
\usepackage{enumitem}
\usepackage{booktabs}
\usepackage{makecell}

\usepackage{tikz}
\usepackage{pgfplots}
\usetikzlibrary {fadings}
\usetikzlibrary{shapes.misc}
\usetikzlibrary{spy}
\usetikzlibrary{angles}
\usetikzlibrary{arrows}
\usetikzlibrary {calc}
\usetikzlibrary{3d}

\tikzset{cross/.style={cross out, draw=black, minimum size=2*(#1-\pgflinewidth), inner sep=0pt, outer sep=0pt},
cross/.default={1.5pt}}

\pgfplotsset{
    legend image with text/.style={
        legend image code/.code={%
            \node[anchor=center, align=center] at (0.3cm,0cm) {#1};
        }
    },
}

\pgfdeclarelayer{fg}    % declare foreground layer
\pgfsetlayers{main,fg}  % set the order of the layers (main is the standard layer)

\def\fwidth{0.5\linewidth}
\def\fheight{0.23\linewidth}
\def\cmwidth{0.35\linewidth}
\def\cmheight{0.35\linewidth}
\def\cbwith{0.7\linewidth}
\def\cbheight{0.37\linewidth}

\def\cdfsysxmax{90}
\def\cdfsyswidth{0.7\textwidth}
\def\xzoompt{0.6}
\def\yzoompt{0.31}
\def\zoomwidth{0.5}
\def\zoomheight{0.9}
\def\zoomxmin{0}
\def\zoomxmax{6}
\def\zoomymin{0.5}
\def\zoomymax{1}
\def\cdffheight{0.35\linewidth}
\def\pmx{4}
\def\pmy{3}
\def\colpscale{0.3}
\def\colpshift{30}

\colorlet{mycolor1}{cyan}%
\colorlet{mycolor2}{magenta}%
\colorlet{mycolor3}{blue}%
\colorlet{mycolor4}{orange}%
\colorlet{mycolor5}{red}%
\colorlet{mycolor6}{violet}%

\usepackage[skip=2pt]{caption}

\newcommand{\phase}[1]{\textit{arg}( #1 )}
\newcommand{\sinc}{\textit{sinc}}
\renewcommand{\vv}[1]{\mathbf{#1}}  % vector notation
\newcommand{\smallsection}[1]{\noindent\textbf{#1}.}

\newcommand{\fig}[1]{Fig.~\ref{#1}}

\newcommand{\secref}[1]{Sec.~\ref{#1}}

\newcommand{\edit}[1]{{\color{black}#1}}

\setcopyright{none}
\begin{document}

%standards
\newacronym{3gpp}{3GPP}{Third Generation Partnership Project}
\newacronym{fr3}{FR3}{Frequency Range 3}
\newacronym{fr2}{FR2}{Frequency Range 2}
\newacronym{fr1}{FR1}{Frequency Range 1}
\newacronym{6g}{6G}{Sixth Generation}
\newacronym{5g}{5G}{Fifth Generation}
\newacronym{4g}{4G}{Fourth Generation}
\newacronym{wifi}{Wi-Fi}{Wireless Fidelity}
\newacronym{cran}{C-RAN}{Cloud Radio Access Network}
\newacronym{pusch}{PUSCH}{Physical Uplink Shared Channel}
\newacronym{dmrs}{DM-RS}{Demodulation Reference Signal}

%general terms
\newacronym{snr}{SNR}{Signal to Noise Ratio}
\newacronym{dpoa}{DPoA}{Differential Phase of Arrival}
\newacronym{tdoa}{TDoA}{Time Difference of Arrival}
\newacronym{ota}{OTA}{Over The Air}
\newacronym{los}{LoS}{Line of Sight}
\newacronym{pdf}{PDF}{Probability Density Function}
\newacronym{ppm}{ppm}{Parts per Million}
\newacronym{cir}{CIR}{Channel Impulse Response}
\newacronym{cfr}{CFR}{Channel Frequency Response}
\newacronym{nlos}{NLoS}{Non Line of Sight}
\newacronym{mmwave}{mmWave}{millimeter wave}
\newacronym{tx}{TX}{Transmitter}
\newacronym{rx}{RX}{Receiver}
\newacronym{ifft}{IFFT}{Inverse Fast Fourier Transform}
\newacronym{ofdm}{OFDM}{Orthogonal Frequency Division Multiplexing}
\newacronym{scs}{ScS}{Subcarrier Spacing}
\newacronym{loess}{LOESS}{LOcally Estimated Scatterplot Smoothing}
\newacronym{cdf}{CDF}{Cumulative Distribution Function}
\newacronym[plural=RBs,firstplural=Resource Blocks (RBs)]{rb}{RB}{Resource Block}
\newacronym{mimo}{MIMO}{Multiple Input Multiple Output}
\newacronym{rms}{RMS}{Root Mean Square}
\newacronym{kf}{KF}{Kalman Filter}
\newacronym{iid}{i.i.d.}{Independent and Identically Distributed}
\newacronym{sota}{SOTA}{State-Of-The-Art}
\newacronym{isac}{ISAC}{Integrated Sensing And Communication}
\newacronym{psd}{PSD}{Power Spectral Density}

%hardware related acronyms
\newacronym{lo}{LO}{Local Oscillator}
\newacronym{if}{IF}{Intermediate Frequency}
\newacronym{rf}{RF}{Radio Frequency}
\newacronym{lna}{LNA}{Low Noise Amplifier}
\newacronym{cfo}{CFO}{Carrier Frequency Offset}
\newacronym{lpf}{LPF}{Low Pass Filter}
\newacronym{hpf}{HPF}{High Pass Filter}
\newacronym{bpf}{BPF}{Band Pass Filter}
\newacronym{p1db}{P1dB}{$1$~dB compression point}
\newacronym{fr4}{FR4}{Flame Retardant 4}
\newacronym[plural=PCBs,firstplural=Printed Circuit Boards (PCBs)]{pcb}{PCB}{Printed Circuit Board}
\newacronym{smd}{SMD}{Surface Mount Device}

%RFSoC
\newacronym[plural=RFSoCs]{rfsoc}{RFSoC}{RF System on a Chip}
\newacronym[plural=NCOs,firstplural=Numerically Controlled Oscillators (NCOs)]{nco}{NCO}{Numerically Controlled Oscillator}
\newacronym{ad}{AD}{Analog to Digital}
\newacronym{da}{DA}{Digital to Analog}

\title{Millimeter-Scale Absolute Carrier Phase-Based Localization in Multi-Band Systems}

\author{Andrea Bedin}
\email{andrea.bedin@imdea.org}
\affiliation{%
  \institution{IDMEA Networks}
  \city{Madrid}
  \country{Spain}
}

\author{Joerg Widmer}
\email{joerg.widmer@imdea.org}
\affiliation{%
  \institution{IDMEA Networks}
  \city{Madrid}
  \country{Spain}
}

\author{Melanny Davila}
\email{10947237@polimi.it}
\affiliation{%
  \institution{Politecnico di Milano}
  \city{Milan}
  \country{Italy}
}

\author{Marco Canil}
\email{marco.canil@imdea.org}
\affiliation{%
  \institution{IDMEA Networks}
  \city{Madrid}
  \country{Spain}
}

\author{Rafael Ruiz}
\email{rafael.ruiz@imdea.org}
\affiliation{%
  \institution{IDMEA Networks}
  \city{Madrid}
  \country{Spain}
}


\begin{abstract}
Localization is a key feature of future \gls{6g} networks with foreseen accuracy requirements down to the millimeter level, to enable novel applications in the fields of telesurgery, high-precision manufacturing, and others.
Currently, such accuracy requirements are only achievable with specialized or highly resource-demanding systems, rendering them impractical for more wide-spread deployment. In this paper, we present the first system that enables low-complexity and low-bandwidth \emph{absolute} 3D localization with \emph{milli\-meter-level accuracy} in generic wireless networks. It performs a carrier phase-based wireless localization refinement of an initial location estimate based on successive location-likelihood optimization across multiple bands. 
Unlike previous phase unwrapping methods, our solution is \emph{one-shot}.
We evaluate its performance collecting ${\sim}350,000$ measurements, showing an improvement of more than \emph{one order of magnitude} over classical localization techniques.
Finally, we will open-source the low-cost, modular FR3 front-end that we developed for the experimental campaign.
\end{abstract}

\maketitle

\section{Introduction}

Localization is a key feature of future \gls{6g} networks \cite{localization_survey_6g, 6gloc_2, 6gloc_3, 6G_XR_healthcare}, with foreseen accuracy requirements at the millimeter and even sub-millimeter level \cite{hexa_localization, sub_cm_requirements} for applications such as precision healthcare and telesurgery \cite{precision_healthcare, 6G_XR_healthcare}, high-precision manufacturing \cite{submm_robot} and augmented reality \cite{localization_survey_6g}. Such millimeter  accuracy requires an improvement of an order of magnitude over current systems, which at best provide accuracies in the centimeter to decimeter range \cite{bounds_1} using complex techniques such as band stitching \cite{wifi_band_stiutching, band_stitching_2} \edit{which is computationally demanding}, large antenna arrays \cite{baseband_cm} \edit{which come with high hardware cost} or ultra-wideband \cite{uwb_loc1, uwb_loc2} \edit{which is less ubiquitous than other technologies and therefore less likely to be already present}. True millimeter-level accuracy has only been achieved with specialized equipment, such as radars localizing modulated backscattering tags \cite{Hawkeye,backscatter_2} and sub-THz systems \cite{localization_100G}. Realizing this with baseband techniques would require extreme bandwidths, which makes such systems very costly \edit{due to the demanding \gls{ad} converter requirements,} and incurs spectrum allocation issues.

The rise of \gls{cran} systems, where coherent processing and a shared local oscillator can be easily implemented \cite{cran_coherent,fiber_lo_distribution}, makes the use of carrier phase information for localization appealing. This is especially true considering the fine range resolution provided by phase measurements at higher frequencies such as the upcoming \gls{fr3} \cite{fr3_1} band.
The adoption of \gls{cran} is also confirmed by industry sources. For example, \cite{Nokia_cran} reports that, "Cloud RAN customers are already running live traffic in their networks", and \cite{Nokia_cran_press, Ericsonn_cran_press} report successful trial deployments. However, up to now, carrier phase location systems have primarily been used for precise \textit{relative} measurements \cite{mtrack,platypus, mmvib}, relying on the well-known strategy of unwrapping the phase to track movement. 
An important drawback of these systems is that they rely on the availability of a continuous recording of the object positions and they stop working whenever tracking is interrupted.
Moreover, since the object position is inferred from the history of past positions, errors accumulate over time.
This precludes applications requiring highly precise \emph{absolute} positioning over long time spans, such as 3D printing with drones \cite{drones_3dprint}, where the drone could move out of coverage to re-load its material supply, resulting in loss of tracking and inability to resume the print with an accurate initial position. 

In this work, we present the first method for wireless communication networks that provides \emph{absolute 3D localization} with a \emph{millimeter-level accuracy}.
Furthermore, unlike previous works \cite{tracking_1, tracking_2, tracking_3, tracking_4, tracking_5}, 
the ultimate accuracy is achieved \emph{one shot}, without filtering trajectories over multiple measurements. This enables tasks that require precise and consistent movement over long periods of time.

To this end, we leverage the concurrent use of carrier phase information across multiple bands. This is a credible solution thanks to the ubiquitous spread of multi-band systems, with devices that support multiple technologies and  inter-band carrier aggregation in \gls{3gpp} standards. For example, modern commercial \gls{4g} networks operate with at least $2$ aggregated carriers in the vast majority of  cases, while \gls{5g}, depending on the operator and location considered, uses carrier aggregation in anywhere between $20$\% and $93$\% of the cases \cite{carrier_aggregation_probability}. While these statistics do not differentiate inter-band and intra-band carrier aggregation, \cite{carrier_aggregation_probability} shows a number of inter-band aggregation cases in the analysis, suggesting that this configuration is commonly used in practice.

In essence, our method provides a robust and effective technique to \emph{enhance a centimeter-accuracy location to millimeter scale} using carrier phase information.
The method is agnostic to how the initial centimeter-level localization is estimated. For example, it could be obtained using current baseband methods. Moreover, in contrast to other approaches that resolve integer ambiguity by combining the observations from many antennas and ensuring the information is consistent \cite{phase_only_localization}, our method only requires $4$ antennas, which is the minimum number required to resolve a 3D location without synchronization between \gls{tx} and \gls{rx}.  This means that in a deployment with, e.g., 20 antenna sites, methods like the one proposed in \cite{phase_only_localization} would need to have \gls{los} for all (or most) antennas, whereas our method can tolerate up to $16$ blocked antennas, which is $80$\% of the total. This mitigates the impact of blockage thanks to the small number of required antennas. Such blockage can be detected and the information from blocked antennas can simply be discarded. 

One key challenge of this technology is related to the breadth of used frequencies: with a diverse set of frequency bands, it is not feasible to have a single antenna covering the whole spectrum. As such, the assumption of having co-located antennas for all bands within the required millimetric accuracy does not hold, even if the antennas are mounted on the same site. 
This precludes us from adopting complexity-reduction techniques from the literature \cite{colocated_antennas,colocated_antennas_2,5g_phase_colocated}, where each antenna can solve a 1-dimensional problem to accurately determine its distance from the target device, and only then aggregating such distance estimates to generate a 3D location. 
Therefore, our method directly fuses the phase information at the 3D location level, resulting in a more flexible design that allows arbitrary antenna deployments. This is also advantageous when considering deployments optimized for communications, where different frequency bands 
may not even use the same cell site due to the different communication ranges.
Thanks to the fact that data fusion is performed at the likelihood level, our method is agnostic to the radio technology used.

While in the paper we focus on \gls{3gpp} systems, as it is the most likely candidate to provide coherent measurements in the near future thanks to \gls{cran}, and it is already routinely using multiple bands \cite{carrier_aggregation_probability}, any other radio technology, if implemented with phase coherence, can be used with this method.

Finally, we evaluate the performance of the method through an extensive measurement campaign, collecting over ${\sim}350,000$ measurements across ${\sim}58,000$ locations in $6$ different multipath environments. These results show that the method can achieve millimetric accuracy in all environments without the need of re-calibration, and is robust to rotations of the \gls{tx} antenna and even relatively high phase errors.
Beyond the innovation brought by the localization technique, with this work we also contribute a cost-effective and modular \gls{fr3} front-end, that has been used in the evaluation of the method and will be made available to the community in an open-source format.

\section{Concepts and feasibility} \label{sec:concepts}

\begin{figure}[t]
\centering
\begin{tikzpicture}
    % Draw the sine wave
    \draw[domain=0:2*pi,samples=100,smooth,thick,blue, line width = 1.5] plot (
        \x,{0.5*sin(deg(2*\x))});
    
    % Draw the horizontal phase ambiguity line
    \draw[dashed,red] (-0.2,0.25) -- (2*pi,0.25);
    \draw[dashed] (17*pi/12,-0.65) -- (17*pi/12,0.65);
    \draw[dashed] (0,-0.65) -- (0,0.65);
    \draw[dashed] (5*pi/12,-0.65) -- (5*pi/12,0.65);
    % Mark the intersections only in the rising direction
    \foreach \x in {5*pi/12, 17*pi/12} {
        \fill[red] (\x,0.25) circle (2.5pt);
        
    }

    \draw[<->](5*pi/12,-0.6) -- node[below] {$\lambda$} (17*pi/12,-0.6);
        \draw[<->](0,-0.6) -- node[above] {$\frac{\phi}{2 \pi} \lambda$} (5*pi/12,-0.6);

    % Labels
    \node[anchor=east] at (-0.2,0.25) {$\phi$};
    \node[below] at (5*pi/12,-0.65) {$k=0$};
    \node[below] at (17*pi/12,-0.65) {$k=1$};
    
    % Axes
    \draw[->] (-0.5,0) -- (2*pi+0.5,0) node[right] {$d$};

\end{tikzpicture}
\caption{Illustration of the phase ambiguity problem.}
\label{fig:phase_amb}
\end{figure}
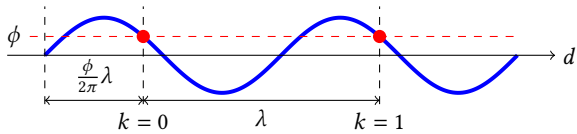

\begin{figure*}[t]
\centering
    \includegraphics{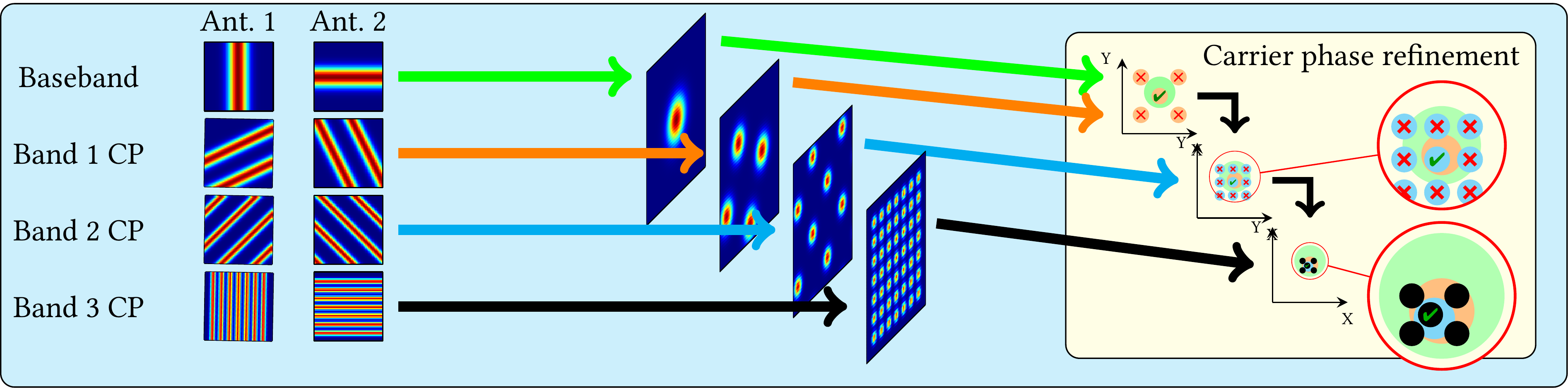}
\vspace{5pt}
\caption{Localization refinement system.}
\label{fig:sysmodel}
\end{figure*}

The key idea behind the proposed method relies on the trade-off between ambiguity and accuracy in carrier phase localization. This ambiguity stems from the fact that the phase can only be measured up to a multiple of $360^\circ$. As a result, a measured phase $\phi$ corresponds to any range of the form $d = (k + \frac{\phi}{2 \pi}) \lambda$, where $k$ is an unknown integer, as illustrated in Fig.~\ref{fig:phase_amb}. Notably, this effect scales with the wavelength $\lambda$, i.e., higher frequencies provide greater accuracy but introduce more potential solutions to the integer ambiguity problem. Solving integer ambiguity with multiband \emph{co-located} antennas is a relatively simple one-dimensional problem, as it only requires determining the distance $d$ between the \gls{tx} and \gls{rx}. To illustrate this, assume we have an initial range estimate of $d = 130~\text{cm} \pm 5~\text{cm}$, and a phase measurement on a low frequency band giving $d = 3~\text{cm} + k_1 \times 10~\text{cm} \pm 0.5~\text{cm}$. It is immediately apparent that $k_1 = 13$ is the only feasible solution, yielding $d = 133~\text{cm} \pm 0.5~\text{cm}$. \edit{Next, assume a higher frequency gives $d = 8~\text{mm} + k_2 \times 1~\text{cm} \pm 1~\text{mm}$. Based on the previously refined value, we know that $k_2 = 132$ and thus $d = 132.8~\text{cm} \pm 1~\text{mm}$}. This demonstrates how, given an accurate enough estimate at each step, the integer ambiguities can be easily resolved in the $1$-dimensional case. In practice though, \edit{perfect antenna co-location} cannot be achieved to the millimeter accuracy simply due to the physical size of the antennas. In addition, antennas are usually positioned to optimize coverage. Thus, the optimal placement differs for each frequency band due to the varying propagation characteristics, which may result in different antenna sites for different bands. Furthermore, as the \gls{tx} has an independent clock and \gls{lo}, and is therefore neither time nor phase synchronized with the \gls{rx}, we cannot measure the absolute value of the range. Thus, the difference between the distances to each antenna needs to be considered.
For these reasons, applying a similar strategy to the $3$-dimensional problem with non-co-located antennas and, therefore, with differences in the corresponding ranges, is not trivial. 

To address this problem, we propose a novel method of combining the information of different bands, where phase information is not merged at the range level, but rather at the 3D location likelihood level. Fig.~\ref{fig:sysmodel} shows examples of the location likelihood that can be obtained by two antennas on the left, and the aggregated likelihood maps in the center.\footnote{For simplicity, absolute phase and time of arrival have been considered whereas, in practice, time and phase difference are used, and the straight bands seen on the left would be hyperbolic.} On the right, we depict how the information from multiple maps is combined, iteratively solving the ambiguity of each frequency band using the information from the respective lower frequencies. In the example on the right of Fig.~\ref{fig:sysmodel}, an initial estimate (represented by the green circle) is used to disambiguate the localization from a low-frequency band (orange circles) in a first refinement step. In the second step, the new estimate from the first band is used to disambiguate a higher frequency estimate (light blue circles). Here we can also see that the initial estimate would not have been accurate enough to disambiguate the second band. The estimate provided by the second band is accurate enough to disambiguate the third one (black circles), resulting in the highly accurate final estimate, which would have also been impossible to disambiguate with only the initial estimate and the first band.  
One natural question is whether the commercially used frequencies have the right numerical properties for this method to work. To assess that, let us consider some of the bands available in modern and future wireless networks. At $1.8$~GHz ($\lambda = 16.6$~cm), phase measurements with an uncertainty of $30^\circ$ (a pessimistic value, compared to the $16.5^\circ$ \emph{worst case} \gls{rms} error measured in our experiments) can provide an accuracy on the order of roughly $\frac{30}{360}\times 16.6\text{cm} = 1.5$~cm, and an uncertainty below $\frac{16.6}{2} = 8.3$~cm from a classic localization system is sufficient to resolve its ambiguity. At $3.5$~GHz ($\lambda = 8.5$~cm) the uncertainty is on the order of $7$~mm, and its $4.25$~cm ambiguity can be comfortably solved with the $1.5$~cm accuracy of the $1.8$~GHz system. Finally, at $10$~GHz (low \gls{fr3}), we can achieve an uncertainty on the order of a few mm, with a $1.5$~cm ambiguity that can be solved with $3.5$~GHz.
Note that the advent of the new \gls{fr3} band facilitates this technology. \edit{ For example, a $57$~GHz \gls{fr2} system in place of the $10$~GHz would require a $0.25$~cm accuracy to solve its ambiguity}, which corresponds to a phase error for the $3.5$~GHz system of $\frac{0.25\text{~cm}}{12\text{~cm}} \times 360^\circ \approx 10^\circ$. While this can be obtained in favorable conditions, it is highly challenging to maintain this performance \textit{reliably} in a realistic multipath environment. In practice, we will show that this method can refine the centimeter accuracy estimates from a classic baseband location system \cite{wifi_band_stiutching, cm_accuracy_rel18} to millimeter scale with just two bands, even under a high multipath. 

With such a high precision system, there are several other aspects to consider. First, a $1$~mm accuracy over a range of $10$~m corresponds to a relative accuracy of $100$ \gls{ppm}. Thus a good quality crystal oscillator, which achieves $10$ \gls{ppm} frequency error  \cite{clock_high_accuracy}, can provide a sufficiently accurate timebase for this application. Further, the thermal expansion coefficient of concrete is on the order of $10 \frac{\text{ppm}}{\text{K}}$, which means that the system can tolerate temperature swings of up to $10$~K before the location of the reference antennas significantly changes due to thermal expansion. This is easily achieved in a climate controlled environment such as a factory or operating room, but may require temperature sensing and compensation in more uncontrolled settings. Another factor that can affect accurate radio measurements are changes in air refractive index (and thus in the propagation speed) due to atmospheric conditions, which are also on the order of a few $10$~s of \gls{ppm} \cite{air_ref_index}. \edit{Finally, at this scale one might be concerned about vibrations of the structure holding the antennas. However, \cite{mmvib} shows that vibrations only introduce displacements lower than $0.1$~mm, which are acceptable for a millimetric accuracy system.} Thus, millimetric radio measurements over a $10$~m range are indeed possible with standard hardware in a reasonably climate-controlled environment.

Finally, we highlight some desirable properties of our method, namely: (a) the \emph{bandwidth requirement is decoupled from the accuracy}, as it only impacts the ability to separate multipath; (b) the accuracy is mainly determined by the highest carrier frequency, and thus is largely independent of bandwidth; (c) the ability of correctly solving the ambiguity is mainly related to the spacing between the carrier frequencies, making it also relatively insensitive to bandwidth. This is highly beneficial for use cases that require precise localization but do not require a large communication bandwidth.
\section{Interferometric multiband location refinement} \label{sec:method}

We now provide a detailed description of the proposed method. In Sec.~\ref{sec:iml} we describe the iterative maximum likelihood algorithm used to solve the multi-dimensional integer ambiguity. In Sec.~\ref{sec:likelihood_calc} we explain how the location likelihood is calculated assuming perfect knowledge of the system impairments. The strategy employed to handle such impairments is elucidated in Sec.~\ref{sec:calibration}, where we discuss the system calibration. Finally, in Sec.~\ref{sec:phase_meas} we provide an intuition on how multipath and phase noise affect phase measurements, and discuss numerical examples to illustrate the order of magnitude of phase impairments we can expect.
 
\begin{figure}[t]
\centering
\begin{subfigure}[t]{\linewidth}
\centering
\begin{tikzpicture}

\begin{axis}[%
width=0,
height=0,
scale only axis,
axis background/.style={fill=white},
legend style={anchor=center, legend cell align=left, align=center, draw=white!15!black,  legend columns=7,font=\small}
]

\addplot[only marks, mark=+, mark options={}, mark size=2.5000pt, draw=black] table[row sep=crcr] {%
0	0\\
0	0\\
};
\addlegendentry{Ground truth}

\addplot[only marks, mark=asterisk, mark options={}, mark size=2.5000pt, draw=orange] table[row sep=crcr] {%
0	0\\
0	0\\
};
\addlegendentry{$\hat{x}_0$}

\addplot[only marks, mark=x, mark options={}, mark size=2.5000pt, draw=red] table[row sep=crcr] {%
0	0\\
0	0\\
};
\addlegendentry{\hspace{-7pt}$\hat{x}_1$}

\addplot[only marks, mark=star, mark options={}, mark size=2.5000pt, draw=blue] table[row sep=crcr] {%
0	0\\
0	0\\
};
\addlegendentry{$\hat{x}_2$}

    \addlegendimage{empty legend}
    \addlegendentry{}

\addplot[color=black,->, line width = 1.5] table[row sep=crcr] {%
0	0\\
0	0\\
};
\addlegendentry{Refinement step}

    \addlegendimage{empty legend}
    \addlegendentry{}

\addplot[color=magenta, line width = 3] table[row sep=crcr] {%
0	0\\
0	0\\
};
\addlegendentry{\makebox[0pt][l]{\gls{fr3} search space}}

    \addlegendimage{empty legend}
    \addlegendentry{}
    
\addplot[color=black, line width = 3] table[row sep=crcr] {%
0	0\\
0	0\\
};
\addlegendentry{\makebox[0pt][l]{FR1 search space}}
    \addlegendimage{empty legend}
    \addlegendentry{}

\end{axis}

\end{tikzpicture}%
\end{subfigure}
 \begin{subfigure}[t]{0.49\linewidth}
        \centering
		\includegraphics{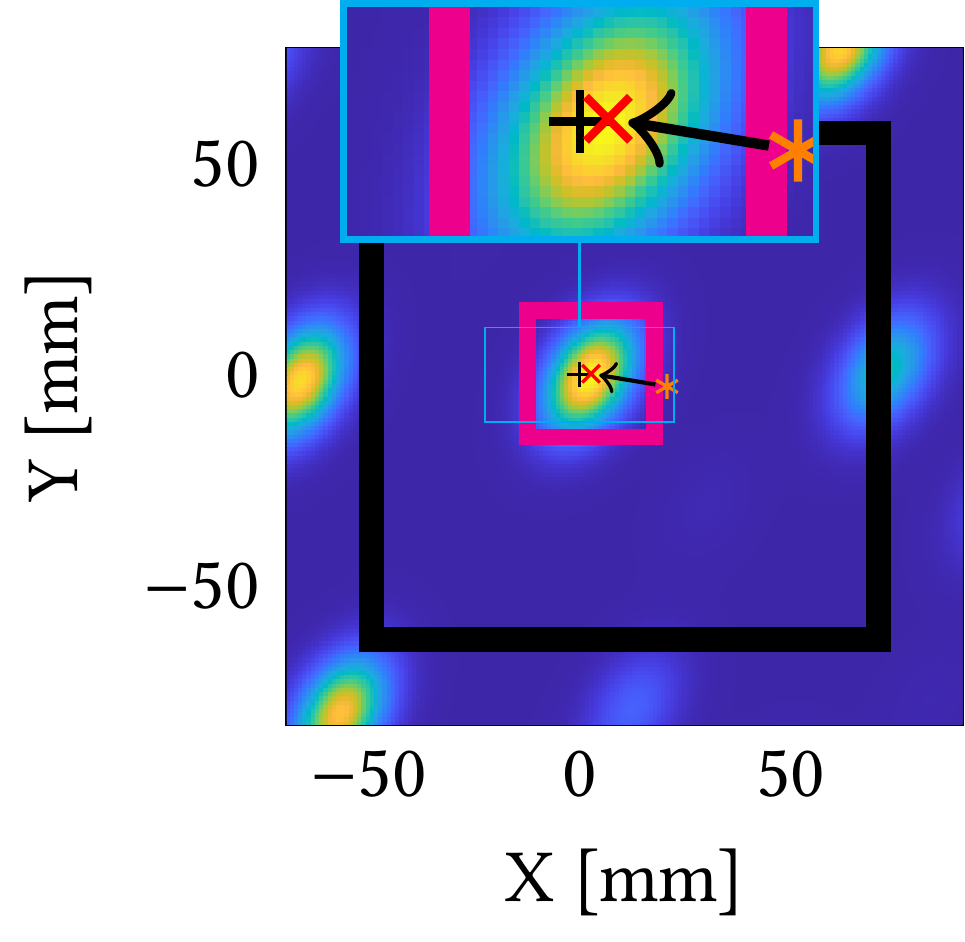}

		\label{fig:likelihood3}
 \end{subfigure}
\begin{subfigure}[t]{0.49\linewidth}
        \centering
		\includegraphics{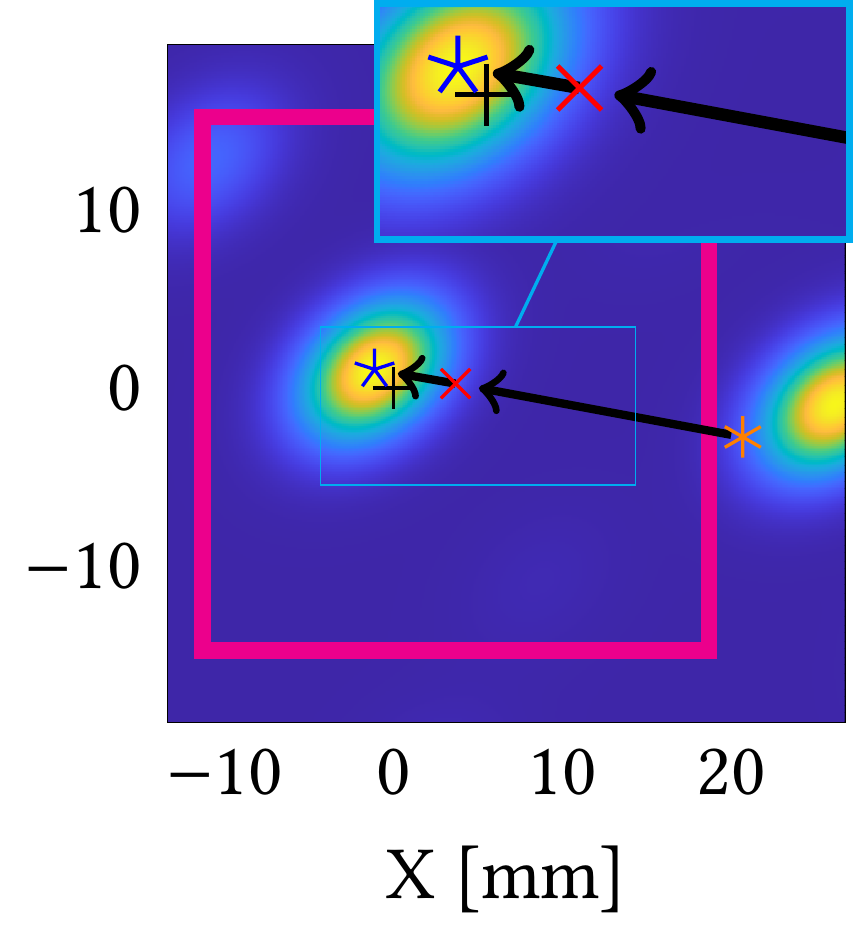}
		\label{fig:likelihood10}
 \end{subfigure}
 %\vspace{-10pt}
\caption{Examples of likelihood maps for $3.25$~GHz (left) and $10.25$~GHz (right).}
\label{fig:likelimap}
\end{figure}
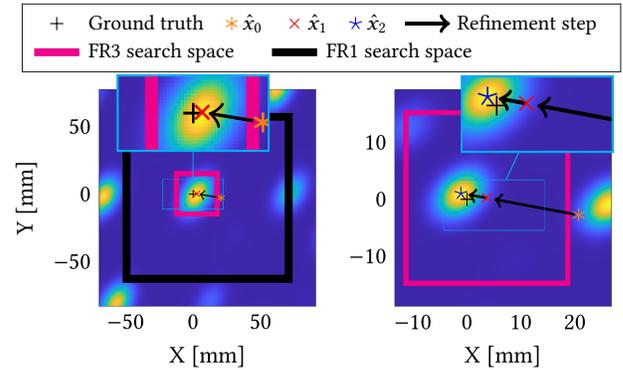

\subsection{Iterative maximum likelihood} \label{sec:iml}
\begin{algorithm}[t]

\small
\caption{Iterative Maximum Likelihood}\label{alg:iml}
\begin{algorithmic}[1]
\State $\hat{x}_0 \gets \text{Initial estimate}$
\For{$k \in \{1,...,K\}$}
\State    $\mathcal{P} \gets \: \textit{peaks} \left( L^{(k)}\left(x \vert \phi^{(k)}\right) \right)$
 \State   $ \hat{x}_{k} \gets  \: p \in \mathcal{P} \textit{ s.t. } \Vert p-\hat{x}_{k-1} \Vert \leq \Vert p'-\hat{x}_{k-1} \Vert \: \forall \: p' \in \mathcal{P}$
\EndFor
\State $\textbf{Return } \hat{x}_{K}$
\end{algorithmic}
\end{algorithm}
Assume we have $K$ systems operating at carrier frequencies $f_k$ ($k \in \{1,...,K\}$, $k < k' \Rightarrow f_k < f_{k'}$). The $k$-th system has $N_k$ \gls{rx} antennas \edit{(or \gls{cran} radios)} spread around the room. The localized target (e.g., a robot) has a single \gls{tx} antenna for each band, located at coordinates $\Delta_k$ with respect to the point on the target we want to localize. We denote the phases measured by antenna $n \leq N_k$ of system $k$ as $\phi^{(k)}_{n}$, and collect the phases of each system $k$ in the vector $\phi^{(k)} = (\phi^{(k)}_1,...,\phi^{(k)}_{N_k})$. With these definitions, we consider the likelihood function $L^{(k)}(x \vert \phi^{(k)})$, $k \leq K$, which is defined as the probability of observing phases $\phi^{(k)}$ given that the \gls{tx} is in position $x$. \edit{An example from real data can be observed in Fig.~\ref{fig:likelimap} (on the left for $3.25$~GHz and right for $10.25$~GHz, the black "+" represents the ground truth location)}. Using this definition, we propose the following method explained in Alg.~\ref{alg:iml}. The method starts from an initial estimate of the location $\hat{x}_0$ \edit{(orange asterisk in Fig.~\ref{fig:likelimap})}, e.g., provided by a baseband location system. The estimate is then refined to $\hat{x}_1$ \edit{(red cross in Fig.~\ref{fig:likelimap})}, defined as the local maximum of $L^{(1)}(x \vert \phi^{(1)})$ which is closest to $\hat{x}_0$. Assuming $\hat{x}_0$ is sufficiently accurate, this leads to a better estimation of the \gls{tx} location. We then re-iterate this refinement using the new estimate as a starting point for the next band, i.e., taking $\hat{x}_k$ as the local maximum of $L^{(k)}(x \vert \phi^{(k)})$ which is closest to $\hat{x}_{k-1}$, until we obtain the final estimate $\hat{x}_K$ \edit{(in our case $\hat{x}_K = \hat{x}_2$ is the blue star in Fig.~\ref{fig:likelimap})}.

The black and magenta boxes in Fig.~\ref{fig:likelimap} represents the search area used to find the likelihood peaks. Given the small search space, this can be done efficiently by exhaustive search\footnote{It takes roughly $20$~ms to localize a target on a laptop with a core i9-13900HX CPU.}. Note that more efficient algorithms exist (e.g., gradient ascent or Newton's method, perhaps with  multiple starting points to ensure the correct peak is identified). However, this is beyond the scope of this paper, as we primarily aim to demonstrate the feasibility of the method. 

\begin{figure}[t]
\centering
 \begin{subfigure}[t]{0.49\linewidth}
        \centering
		\includegraphics{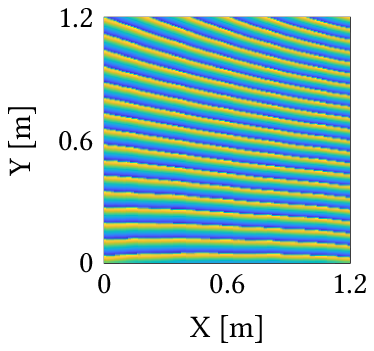}
        \caption{$3.25$~GHz.}
		\label{fig:phase_3}
 \end{subfigure}
\begin{subfigure}[t]{0.49\linewidth}
        \centering
		\includegraphics{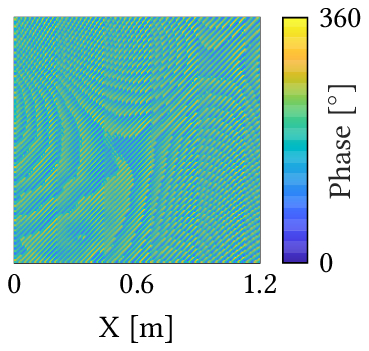}
        \caption{$10.25$~GHz.}
		\label{fig:phase_10}
 \end{subfigure}
\vspace{10pt}
\caption{Examples of the measured phase differences.}
\label{fig:phase_example}
\end{figure}

\subsection{Likelihood calculation} \label{sec:likelihood_calc}

Assuming we have a (possibly imperfect) knowledge of the position $a^{(k)}_n$ of the $n$-th \gls{rx} antenna of the $k$-th system, and $x_k = x + \Delta_k$ is the 3D location of the \gls{tx} antenna for band $k$ when the target is at location $x$, we model the measured phase for each antenna as
\begin{equation}
\displaystyle  \theta^{(k)}_n(x) = \text{mod} \bigg(  \frac{2 \pi f_k \Vert a^{(k)}_n - x_k \Vert}{c} +  \varphi^{(k)} + \hat{\gamma}^{(k)}_n(x_k) + w(\sigma), 2 \pi \bigg),
\end{equation}
where: The first term is the basic model of the phase based on the distance between the \gls{tx} and \gls{rx}. The term $\varphi^{(k)}$ represents a phase offset that is constant across all antennas for a specific band, but can vary over time. This term is due to the lack of synchronization between the \gls{tx} and \gls{rx}, both in the sampling time and in the \gls{lo} phase changes induced by the \gls{cfo}.  The term $\hat{\gamma}^{(k)}_n(x)$ is a calibration function that encodes all deviations from the simple model due to the imperfect knowledge of antenna placement, the variation in the phase center of the antennas, and other factors. The details of how this function is obtained are discussed in Sec.~\ref{sec:calibration}. Noise $w(\sigma)$ is a Gaussian random variable of standard deviation $\sigma$, independent between antennas and bands.
Note that the \gls{tx} antenna displacement $\Delta_k$ depends on the orientation of the device, which we assume is provided by external means such as radio measurements or magnetometers (further discussed in Sec.~\ref{sec:results_orientation} and Sec.~\ref{sec:results_tdoa}). Moreover, we assume $\Delta_k = 0$ for $k=K$, i.e., the final estimate reflects the location of the phase center for the \gls{tx} antenna operating at the highest frequency band. As the phase difference $\varphi^{(k)}$ is unknown and time varying, it cannot be calibrated. Thus, it is well known that phase differences should be considered to eliminate its effect. In particular, given a reference antenna (assume antenna $N_k$ without loss of generality), we consider the phase difference $\Omega^{(k)}_n(x)$, $n \leq N_k-1$ defined as $\Omega^{(k)}_n(x) = \theta^{(k)}_n(x) - \theta^{(k)}_{N_k}(x)$, which applying the addition property of the modulo function results in
\begin{align}
\nonumber \Omega^{(k)}_n(x) = \text{mod} \bigg( & \frac{2 \pi f_k}{c}\left( \Vert a^{(k)}_n - x_k \Vert - \Vert a^{(k)}_{N_k} - x_k \Vert \right) + \\
&\hat{\gamma}^{(k)}_n(x_k)  - \hat{\gamma}^{(k)}_{N_k}(x_k) + w, 2 \pi \bigg)
\end{align}
Examples of the measured $\Omega^{(k)}_n(x)$ can be seen in Fig.~\ref{fig:phase_example}. As expected, the $3.25$~GHz phase shown in Fig.~\ref{fig:phase_3} exhibits relatively slow variations and wraps around every ~$100$~mm. In contrast, the phase at $10.25$~GHz in Fig.~\ref{fig:phase_10} exhibits much faster variations and wraps around roughly every $30$~mm\footnote{The large scale features shown in Fig.~\ref{fig:phase_10} are moiré patterns due to the ratio between measurement grid spacing and the wavelength.}. With these definitions and the knowledge of $\gamma^{(k)}_n(x_k) = \hat{\gamma}^{(k)}_n(x_k)  - \hat{\gamma}^{(k)}_{N_k}(x_k)$, that will be discussed in Sec.~\ref{sec:calibration}, we can write the location likelihood function
\begin{equation}
L^{(k)}(x,\phi^{(k)})= \prod_{n=1}^{N_k-1} \sum_{i=-\infty}^{\infty} \psi\left( \Omega^{(k)}_n(x) - (\phi^{(k)}_{n} - \phi^{(k)}_{N_k}) + 2 \pi i, \sigma \right),
\end{equation}
where $\psi(\cdot, \sigma)$ is the \gls{pdf} of the normal distribution with standard deviation $\sigma$, and the infinite sum reflects the periodicity of the modulo function. Note that the infinite sum has only few non-negligible terms, thus making it computable.

\subsection{System calibration} \label{sec:calibration}
\begin{figure}[t]
    \centering
    \begin{minipage}{0.49\linewidth}
        \centering
        \includegraphics[width=\linewidth]{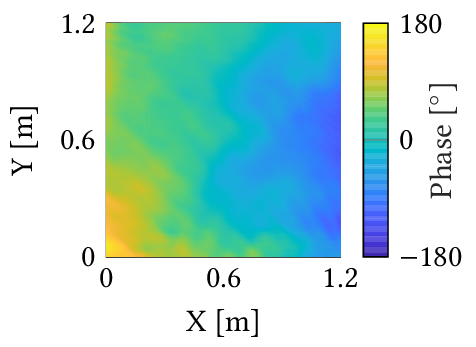}
        \caption{Calibration function example.}
        \label{fig:phase_diff}
    \end{minipage}
    \hfill
    \begin{minipage}{0.49\linewidth}
        \centering
        \def\axwidth{0.95\linewidth}
\def\axheight{0.7 * \linewidth}
\def\radius{0.1 * \linewidth}

\def\scale{0.75}

\begin{tikzpicture}

\coordinate (correct_end) at (0.8*\scale*\axwidth, 0.4*\scale*\axwidth);
\coordinate (multipath_end) at (0.7*\scale*\axwidth, 0.6*\scale*\axwidth);
\coordinate  (O) at (0,0);
\coordinate  (X) at (\axwidth,0);

\draw[->, line width = 2] (-0.1*\scale*\axwidth, 0) -- (\axwidth*\scale, 0)  node [below] {Re$(h)$};
\draw[->, line width = 2] (0,-0.1*\scale*\axwidth) -- (0,\axheight*\scale)  node [right] {Im$(h)$};
\draw[-stealth, line width = 1.5, mycolor1]  (0,0) -- node[below, xshift=6.5mm, sloped] {$h_{LoS}$} (correct_end) ;
\draw[-stealth, line width = 1.5, mycolor2] (correct_end) --  node[above, sloped] {$h_{NLoS}$} (multipath_end);
\draw[-stealth, line width = 1.5, mycolor1] (0,0) --  node[below, sloped, above] {$h$} (multipath_end);
  \pic[pic text=$\Delta \phi$, mycolor2, draw, angle radius=1.2cm, angle eccentricity = 1.2] {angle = correct_end--O--multipath_end};
  \pic[pic text=$\phi$, mycolor1, draw, angle radius=1cm, angle eccentricity = 1.2] {angle = X--O--correct_end};
  \pic[draw,angle radius=0.25cm] {right angle = O--correct_end--multipath_end};
\end{tikzpicture}%
        \caption{Contribution of the NLoS component to the phase.}
        \label{fig:phasors}
    \end{minipage}
\end{figure}

As mentioned above, we encode the \gls{rx} antennas' location error, phase center variation, cable delay, and any other systematic impairments into the calibration function $\gamma^{(k)}_n(x_k)$. Given an accurate enough location estimate of the \gls{rx} antennas (within a few cm), we found that the function $\gamma^{(k)}_n(x_k)$ is quite smooth, with peak-to-peak values on the order of a few hundreds of degrees across the whole area, as exemplified in Fig.~\ref{fig:phase_diff} which shows its measured value for a $10.25$~GHz system.  This makes it possible to fit the function with a limited number of datapoints, which can be collected by manually measuring the location of the \gls{tx}.

Thus, to estimate $\gamma^{(k)}_n(x_k)$ we require the measurement of the phase difference in a small number (a few tens) of known locations. In practice, this has to be achieved by accurately measuring the location of the \gls{tx} with non-radio techniques. While we recognize this task requires a significant effort and expensive equipment, it needs to be performed only once, for a limited number of points, and it can be provided as a service at the time of installing the system. In our opinion, given the exceptional accuracy provided by the system, this makes the effort acceptable. Furthermore, as highlighted in Sec.~\ref{sec:discussion}, self-calibration methods are possible and will be explored in future work. To determine the function $\gamma^{(k)}_n(x_k)$, we first take the calibration points and perform a $2D$ phase unwrapping\footnote{All 2-dimensional phase unwrapping has been realized using \cite{unwrap_code}.} to avoid discontinuities in the function that would make fitting more difficult. Further, we remove the mean value (which 
%corresponds to an offset that is constant over position, and 
is introduced by the difference in the cable length between the antennas) from the unwrapped calibration points. The unwrapped points are then fitted with a \gls{loess} approximation, which, after re-introducing the mean offset is used as the estimate of $\gamma^{(k)}_n(x_k)$.

\subsection{OTA phase measurement considerations} \label{sec:phase_meas}

\edit{Obtaining } accurate \gls{ota} phase measurements \edit{might be a concern} due to multipath and phase noise. However, with the introduction of the \gls{fr3} band, the stringent phase accuracy requirements to resolve the ambiguity can be relaxed, enabling a practical and robust localization even in the presence of phase errors as large as ${\sim}30^\circ$ (see Fig.~\ref{fig:cdf_vs_phaseerr}).

\textbf{Multipath considerations} While we experimentally prove that accurate localization in the presence of multipath is practically achievable in Sec.~\ref{sec:results}, in this section we provide an intuitive explanation of why this is the case.
Let $h(t)$ be the \gls{cir} of the measured channel, and $\tau$ the delay associated with the \gls{los} component. The \gls{cir} at $\tau$ can be written as $h(\tau) = h_{LoS}(\tau) + h_{NLoS}(\tau)$, 
where $h_{LoS}(\tau)$ corresponds to the \gls{los} component of which we aim to measure the phase, while $h_{NLoS}(\tau)$ accounts for multipath. As illustrated in Fig.~\ref{fig:phasors}, the worst-case error occurs when $\phase{ h_{LoS}(\tau)} - \phase{h_{NLoS}(\tau)} \edit{\approx} \pm \frac{\pi}{2}$ \edit{(for small $h_{NLoS}(\tau)$)}, bounding the phase error as:
\begin{equation}  
\vert \Delta \phi \vert  = \vert \phase{ h_{LoS}(\tau)} - \phase{ h(\tau)} \vert \leq \tan^{-1}\left( \frac{\left\vert h_{NLoS}(\tau)\right\vert}{\left\vert h_{LoS}(\tau)\right\vert}  \right).  
\end{equation}  

Considering a channel with bandwidth $B$ and a single dominant multipath component with a delay $\tau' > \tau$ and a relative amplitude $\alpha$ with respect to the \gls{los}, its impact on the \gls{los}, using $1/x$ as an upper bound of the sinc function, satisfies
\begin{equation}  
\frac{\left\vert h_{NLoS}(\tau)\right\vert}{\left\vert h_{LoS}(\tau)\right\vert} = \left\vert \frac{\sinc \left( B (\tau' - \tau) \right)}{\alpha }  \right\vert \leq \frac{1}{ \alpha  B (\tau' - \tau)},  
\end{equation}  
which yields the bound $\vert \Delta \phi \vert   \leq \tan^{-1}\left( \frac{1}{ \alpha  B (\tau' - \tau)}  \right).  $

As an example, consider a reflector with a reflection coefficient of $0.5$ ($\alpha = 2$), producing a path with an excess distance of $3$~m ($\tau' - \tau = 10$~ns), and a bandwidth of $100$~MHz. Applying the above bound $
\vert \Delta \phi \vert \leq \tan^{-1}\left( \frac{1}{ 2 \times  100\times 10^6  \times 10 \times 10^-9}  \right) \approx 26.5^\circ. $

\edit{Note} that a reflection coefficient of $0.5$ is quite high. \edit{For example, by generating $100$~k realizations of the \gls{3gpp} TDL10D channel model \cite{3gppchmodel1, 3gppchmodel2, TDL30Dmatlab} we observed a ratio $\geq 0.5$ between the gain of the largest tap and the \gls{los} in only $0.1$\% of cases. Further, to achieve the bound the phase difference between \gls{los} and \gls{nlos} component needs to be  close to $90^\circ$.} Thus, we can expect a naive method, such as using the phase associated with the \gls{cir} amplitude peak, to provide sufficiently accurate phase estimates. While this analysis and example are not intended as a proof that the multipath induced phase impairment is sufficiently small, they do provide an intuitive explanation of the reason behind this observation. The multipath resilience of the method is substantiated by the experimental results in Sec.~\ref{sec:results}, and in particular by the fact that the system calibration we perform in the absence of additional reflectors still applies to a rich multipath environment while maintaining millimetric accuracy.

\textbf{Phase noise} Another effect that is known to impact phase measurements is phase noise. To evaluate its impact, we simulated a phase noise impaired channel using MATLAB's implementation \cite{phase_noise_matlab}. To define realistic phase noise performance, we selected a mid-range commercial PLL \cite{pll_phase_noise}. The phase noise of this device, which is reported as the \gls{psd} of the output signal at different offsets from the generated tone, is of  $-107\frac{\text{dBc}}{\text{Hz}}$, $-127\frac{\text{dBc}}{\text{Hz}}$, $-129\frac{\text{dBc}}{\text{Hz}}$ and $-147\frac{\text{dBc}}{\text{Hz}}$ at $100$~kHz, $800$~kHz, $1$~MHz and $10$~MHz offset, respectively. To ensure a conservative estimate of the phase error, we increase the phase noise \gls{psd} by $20$~dB at all frequency offsets. We then performed the phase extraction as described in Sec.~\ref{sec:platform} (using a perfect time and frequency synchronization between \gls{tx} and \gls{rx} to isolate the impact of phase noise) on such signal. In these simulations, we found that the phase noise induced error in the phase measurement has a standard deviation of $2.6^\circ$, which is negligible compared to the required accuracy and the multipath impact.

\section{System Platform}
\label{sec:platform}

\subsection{Implementation Overview}

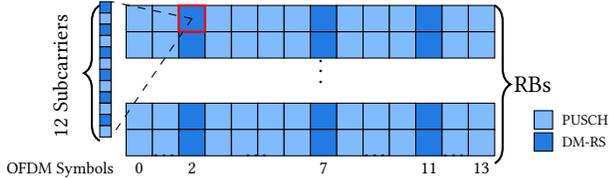
\begin{figure}[t]
\vspace{-0.5cm}
\centering
\begin{tikzpicture}
    \definecolor{PUSCHColor}{RGB}{129,187, 255}    % Blue
    \definecolor{DMRSColor}{RGB}{33, 123, 226} % Skyblue
    
    % Main matrix size
    \def\rows{2}
    \def\cols{14}
    \def\cellsize{0.35} % Cell size

    \foreach \x in {1,...,\cols} {
        \foreach \y in {1,...,\rows} {
            %Columns 3, 8, and 12
            \ifnum \x=3
                \filldraw[DMRSColor, draw=black] (\x*\cellsize,-\y*\cellsize) rectangle ++(\cellsize,-\cellsize);
            \else
            \ifnum \x=8
                \filldraw[DMRSColor, draw=black] (\x*\cellsize,-\y*\cellsize) rectangle ++(\cellsize,-\cellsize);
            \else
            \ifnum \x=12
                \filldraw[DMRSColor, draw=black] (\x*\cellsize,-\y*\cellsize) rectangle ++(\cellsize,-\cellsize);
            \else
                \filldraw[PUSCHColor, draw=black] (\x*\cellsize,-\y*\cellsize) rectangle ++(\cellsize,-\cellsize);
            \fi
            \fi
            \fi
        }
    }

    \node at (0.6*\cols*\cellsize, -0.5\rows*\cellsize-0.47*\rows) {\vdots};

    \foreach \x in {1,...,\cols} {
        \foreach \y in {1,...,\rows} {
            % Columns 3, 8, and 12
            \ifnum \x=3
                \filldraw[DMRSColor, draw=black] (\x*\cellsize,-\y*\cellsize-\rows*\cellsize-0.6) rectangle ++(\cellsize,-\cellsize);
            \else
            \ifnum \x=8
                \filldraw[DMRSColor, draw=black] (\x*\cellsize,-\y*\cellsize-\rows*\cellsize-0.6) rectangle ++(\cellsize,-\cellsize);
            \else
            \ifnum \x=12
                \filldraw[DMRSColor, draw=black] (\x*\cellsize,-\y*\cellsize-\rows*\cellsize-0.6) rectangle ++(\cellsize,-\cellsize);
            \else
                \filldraw[PUSCHColor, draw=black] (\x*\cellsize,-\y*\cellsize-\rows*\cellsize-0.6) rectangle ++(\cellsize,-\cellsize);
            \fi
            \fi
            \fi
        }
    }
    
    % Zoom cell coordinates
    \def\zoomrow{1}
    \def\zoomcol{3}

    % Red square around the zoomed cell
    \draw[red, thick] (\zoomcol*\cellsize,-\zoomrow*\cellsize) rectangle ++(\cellsize,-\cellsize);

    % Zoom cell
    \def\zoomXOffset{8cm} % X movement of the zoomed cell
    \def\zoomYOffset{-0.5cm} % Y movement of the zoomed cell
    \def\zoomCellSize{0.15} % Zoomed cell size

    % Calculate the zoom origin outside the main grid
    \node[anchor=west] (zoom_origin) at (0,0){};%(\cols*\cellsize, -\rows * \cellsize - \zoomYOffset) {};

    %Origin coordinates
    \coordinate (zoom_origin) at (0,-0.3);

    \def\subc{12}

    % Zoomed RB with PUSCH and DMRS slots
    \foreach \i in {0,...,11} {
        \ifodd\i
            \filldraw[PUSCHColor, draw=black] (zoom_origin) ++(0,-\i*\zoomCellSize) rectangle ++(\zoomCellSize,-\zoomCellSize);
        \else
            \filldraw[DMRSColor, draw=black] (zoom_origin) ++(0,-\i*\zoomCellSize) rectangle ++(\zoomCellSize,-\zoomCellSize);
        \fi
    }

    % Dashed line
    \draw[dashed] 
        (\zoomcol*\cellsize + \cellsize/2,-\zoomrow*\cellsize - \cellsize/2) 
        -- (\zoomCellSize,-0.3);

    \draw[dashed] (\zoomcol*\cellsize + \cellsize/2,-\zoomrow*\cellsize - \cellsize/2) -- (\zoomCellSize,-0.3-\subc*\zoomCellSize);

    % Resource Blocks Label (Above the main grid)
    % \node[right, rotate=270] at (\cols/2+\cellsize/2, -\cellsize) {\glspl{rb}};

    % 12 Subcarriers Label (Right of the Zoomed-in Column, rotated)
    %\node[right, rotate=90] at (-\cellsize,-\subc*\zoomCellSize-2*\zoomCellSize) {\footnotesize12 Subcarriers};

    \draw [decorate,decoration={brace,amplitude=6pt,mirror},thick] 
    (0, -\cellsize) -- (0, -\subc*\zoomCellSize-\cellsize) 
    node[at start,left=15pt, rotate=90] {\small12 Subcarriers};

     % Symbol Numbers (Below columns 1 and 14) - Added font size command
    \node[below, scale=0.7, transform shape] at (-3*\cellsize/2, -\rows-\cellsize) {OFDM Symbols};
    \node[below, scale=0.75, transform shape] at (3*\cellsize/2, -\rows-\cellsize) {0};
    \node at (5*\cellsize/2, -\rows-\cellsize) {\dots};
    \node[below, scale=0.75, transform shape] at (7*\cellsize/2, -\rows- \cellsize) {2};
    \node at (12*\cellsize/2, -\rows- \cellsize) {\dots};
    \node[below, scale=0.75, transform shape] at (17*\cellsize/2, -\rows- \cellsize) {7};
    \node at (21*\cellsize/2, -\rows-\cellsize) {\dots};
    \node[below, scale=0.75, transform shape] at (25*\cellsize/2, -\rows- \cellsize) {11};
    \node at (27*\cellsize/2, -\rows-\cellsize) {\dots};
    \node[below, scale=0.75, transform shape] at (\cols*\cellsize+\cellsize/2, -\rows- \cellsize) {13};

   %Legends
    \def\legendXOffset{2.5*\cellsize} % Horizontal offset for the legend
    \def\legendYStart{-2.5*\rows*\cellsize} % Starting vertical position for the legend
    \def\legendBoxSize{0.7*\cellsize} % Size of the colored boxes in the legend
    \def\legendTextSize{\small}

    % PUSCH square
\filldraw[PUSCHColor, draw=black] (\cols*\cellsize + \legendXOffset, \legendYStart) 
    rectangle ++(\legendBoxSize, -\legendBoxSize);
\node[anchor=west, scale=0.7] at (\cols*\cellsize + \legendXOffset + 1.2*\legendBoxSize, \legendYStart - 0.5*\legendBoxSize) 
    {\legendTextSize PUSCH};

% DM-RS square
\filldraw[DMRSColor, draw=black] (\cols*\cellsize + \legendXOffset, \legendYStart - 1.2*\legendBoxSize) 
    rectangle ++(\legendBoxSize, -\legendBoxSize);
\node[anchor=west, scale=0.7] at (\cols*\cellsize + \legendXOffset + 1.2*\legendBoxSize, \legendYStart - 1.2*\legendBoxSize - 0.5*\legendBoxSize) 
    {\legendTextSize DM-RS};

    % RBs curly bracket
    \draw [decorate,decoration={brace,amplitude=6pt},thick] 
    (\cols*\cellsize+\cellsize, -\cellsize) -- (\cols*\cellsize+\cellsize, -2*\rows*\cellsize-3*\cellsize) 
    node[midway,right=4pt] {\glspl{rb}};

\end{tikzpicture}
\caption{Structure of the \gls{5g} slot.}
\label{fig:grid}

\end{figure}

For the evaluation, we use a baseband processor based on \cite{MIMORPH}, which uses an \gls{rfsoc} ZCU208 board \cite{RFSOCZCU208} integrating Giga-sampling rate \gls{ad} and \gls{da} converters, logic elements, DDR memory and 10GbE interfaces. This setup supports $8$x$8$ coherent and synchronized \gls{mimo} with up to $400$~MHz of bandwidth, and utilizes up/down sampling filters to operate in different bandwidth modes. \glspl{rfsoc} also include \glspl{nco} embedded in the \gls{ad}/\gls{da} converters, enabling independent center frequency adjustment for the up/down conversion without losing synchronization between converters. This feature allows direct conversion to the \gls{rf} and \gls{if} domains for the \gls{fr1} and \gls{fr3} bands respectively. We leverage the memory based design of \cite{MIMORPH} to transmit a \gls{5g} Release 18 standard compliant waveform and store the received IQ samples for post-processing in MATLAB.
In particular, the signal is an uplink waveform where all available \glspl{rb} are allocated to the \gls{pusch}, the physical layer channel designated for uplink data transmission. Alongside with the data, the \gls{pusch} comprises some \gls{dmrs} symbols used for channel estimation. \gls{5g} offers great flexibility on the allocation of the \gls{dmrs}. Namely, it specifies one mandatory and up to three additional symbols to be used for \gls{dmrs} \cite{release18DMRS}. In our experiment, we allocate a total of three \gls{dmrs} positions to obtain high quality channel estimates. As the current \gls{5g} standard does not include any specifications for \gls{fr3}, we selected the parameters for the $10.25$~GHz carrier based on the available ranges for \gls{fr1} and \gls{fr2} numerologies. In particular, \gls{5g} specifies a range for \gls{scs} of $15$~kHz to $60$~kHz for \gls{fr1} and $60$~kHz to $960$~kHz for \gls{fr2}, while the channel bandwidth ranges from $3$~MHz to $100$~MHz for \gls{fr1} and $50$~MHz up to $2$~GHz for \gls{fr2} \cite{release18Part1, release18Part2}. Based on these values, we select a $240$~kHz \gls{scs} and a $400$~MHz bandwidth for our hypothetical \gls{fr3} signal. The detailed configuration parameters are given in Tab.~\ref{tab:freq}, and the resulting resource grid is depicted in Fig. \ref{fig:grid}, where we can see in blue the \gls{dmrs} symbols and in sky-blue the user's data.

\begin{table}[t]
  \caption{Configuration Parameters}
  \label{tab:freq}
  \begin{tabular}{lcc}
    \toprule
    \multicolumn{1}{l}{ }& \gls{fr1}& \gls{fr3}\\
    \midrule
    \gls{scs} & 30 kHz & 240 kHz\\   
    \glspl{rb} & 133 & 139\\
    \gls{pusch} Mapping& \multicolumn{2}{c}{Type A}\\
    \gls{dmrs} Configuration & \multicolumn{2}{c}{Type 1}\\
    \gls{dmrs} Positions & \multicolumn{2}{c}{2,7,11}\\   
  \bottomrule
\end{tabular}

\end{table}

The \gls{rfsoc} is configured to utilize two \gls{tx} and eight \gls{rx} channels. One \gls{tx} operates in \gls{fr1}, using the \gls{nco} to perform direct conversion to \gls{rf} at $3.25$~GHz. The second \gls{tx} transmits the \gls{fr3} \gls{if} signal at the same frequency. The \gls{if} signal is then mixed with a $7$~GHz \gls{lo}, generated by a dual-channel synthesizer \cite{synth}, using the \gls{fr3} upconverter presented in Sec.~\ref{sec:frontends}, to obtain a $10.25$~GHz carrier. Accordingly, four receivers operate directly in the \gls{rf} domain at \gls{fr1}, while the remaining four operate at a $750$~MHz \gls{if} frequency to receive the \gls{fr3} signal, which is generated by mixing the $10.25$~GHz \gls{rf} signal with a $9.5$~GHz \gls{lo} using the downconverters presented in Sec.~\ref{sec:frontends}. The  \gls{rx} \gls{lo} is generated by the second channel of the synthesizer \cite{synth}, and distributed through the \gls{lo} distribution board to guarantee coherence between the \glspl{rx}. Note that the use of different \gls{lo} frequencies at the \gls{tx} and \gls{rx}, while motivated by limiting the \gls{lo} leakage at the \gls{tx}, also ensures the non-coherence between \gls{rx} and \gls{tx} \glspl{lo}, making the experiment realistic. The received signal is processed using MATLAB's \gls{5g} toolbox \cite{5gtoolbox} to extract the \gls{cfr}. To obtain the \gls{cir} estimate, we then apply an \gls{ifft} with an upsampling factor equal to 32. The \gls{los} component is then identified by finding the location of the earliest peak in the \gls{cir}, and the measured phase is simply taken as the argument of its complex amplitude. As the channel estimate for the data symbols is just an interpolation of the \gls{dmrs} symbols, to reduce the computational complexity we only compute the \gls{cir} for the $3$ symbols allocated for \gls{dmrs}. The phase difference with the reference antenna is calculated for each symbol individually, and its value is then averaged over the three \gls{dmrs}.\footnote{Using \cite{meanangle} to avoid phase wrapping issues.}  The \gls{rx} antennas are placed at the corners of a gantry that can move the \gls{tx} antenna module with its $3.25$~GHz and $10.25$~GHz omnidirectional antennas anywhere in a $1.2$~m$\times 1.2$~m $2$-dimensional area with high accuracy. \edit{We note that the gantry is made of alluminum, with a thermal expansion coefficient of $23\frac{\text{ppm}}{\text{K}}$, so even over a large temperature change of $30$~K, the structure will expand by less than $23\frac{\text{ppm}}{\text{K}} \times 30\text{~K} \times 1.2\text{~m} = 0.8\text{~mm}$, maintaining a millimetric accuracy of the ground truth.} This area is limited by the size of the gantry itself, that is needed to obtain the highly accurate ground truth. It does not imply a limitation of the method itself, which is applicable to an arbitrarily large area given a sufficient \gls{snr} \edit{and a well controlled temperature}. Further, we note that the phase measurement accuracy is mostly limited by multipath, thus we do not expect the \gls{snr} degradation due to a larger distance to impact the localization performance significantly. In contrast, a larger distance might lead to a larger difference in delay between the \gls{los} and reflected paths, thus mitigating the multipath issue.

\begin{figure}[t]
\centering
\input{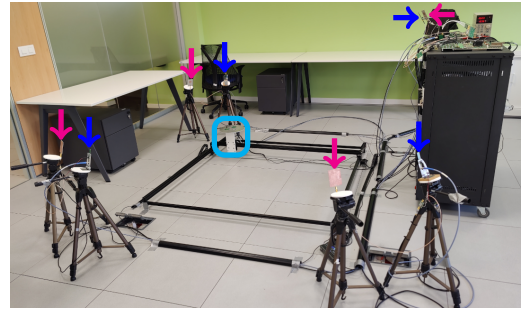}
\caption{Measurement setup. The movable \gls{tx} antenna module is highlighted in cyan, the $3.25$~GHz \gls{rx} antennas in magenta and the $10.25$~GHz \gls{rx} antennas in blue.}
\label{fig:physical_setup}
\end{figure}

\begin{figure}[t]
\centering
\input{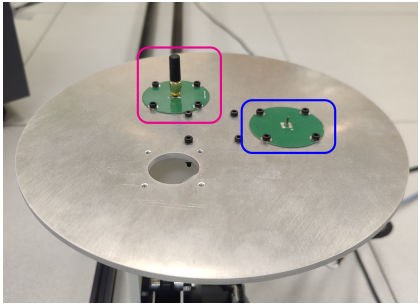}
\caption{Transmitter antenna module. The $3.25$~GHz antenna is highlighted in magenta and the $10.25$~GHz \edit{one} in blue.}
\label{fig:TX_antenna}
\end{figure}

The \gls{tx} antenna module, shown in Fig.~\ref{fig:TX_antenna}, comprises a  circular mounting base capable of housing up to three antennas. In this setup, we installed a commercial $3.25$~GHz whip antenna and a custom-made $10.25$GHz monopole. These antennas were chosen as a relatively omnidirectional radiation pattern is required, since the \gls{rx} antennas are placed all around the \gls{tx}. Further, the antennas and the mounting base are designed to be circularly symmetric. This is done to minimize the center of phase variations when rotating the antenna module. \edit{When rotating the antenna around its phase center, we measured a phase variation of $4.35^\circ$ RMS. For practical deployments, more sophisticated antenna designs can be built with stable phase center \cite{PCV_spiral, PCV_ant_GNSS, PCV_ant_horn}.}

Since the setup is not capable of moving the \gls{tx} in the vertical direction, the evaluation is performed in $2$D, using coordinates of the form $(x,y,0)$ for the \gls{tx} location, while maintaining a $3$D location for the \gls{rx} antennas, since they are placed above the plane. The experimental setup is shown in Fig.~\ref{fig:physical_setup}. The \gls{tx} antenna is moved to a grid of locations, and at each location we receive $3$ \gls{5g} slots, which are independently processed and treated as three separate measurements, to make the dataset more redundant and increase our statistical confidence. As the placement of the \gls{tx} is done with sub-millimeter precision by the gantry, we use the location provided by its stepper motors as ground truth for the performance evaluation. We note that, despite the fact that for simplicity we use a single baseband processor for the experiment, we  neither  exploit the time nor the phase synchronization between \gls{rx} and \gls{tx} that such a system could provide. We ensure this by collecting and processing the data for each band independently, one after the other, as well as performing cyclic prefix-based time and frequency synchronization at the \gls{rx}, which introduce a realistic residual \gls{cfo} and timing offset due to the imperfect estimation.

\subsection{FR3 distributed array} \label{sec:frontends}

\begin{figure}[t]
\centering
 \begin{subfigure}[t]{0.45\linewidth}
        \centering
		\input{figures/hardware/downconverter}
        \caption{Downconverter.}
		\label{fig:dc}
 \end{subfigure}
\begin{subfigure}[t]{0.45\linewidth}
        \centering
		\input{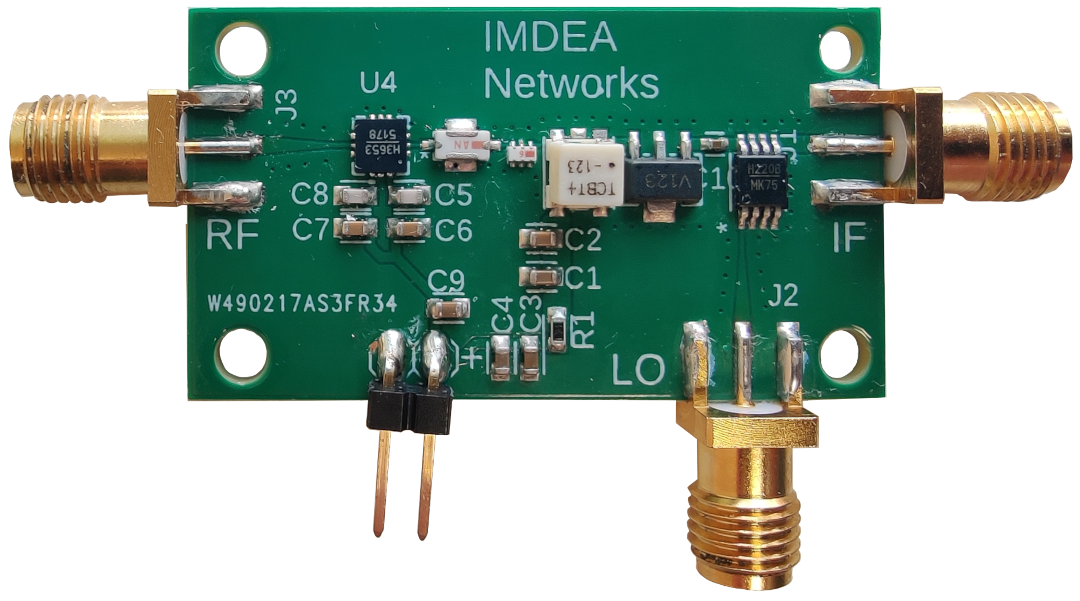}
        \caption{Upconverter.}
         \vspace{10pt}
		\label{fig:uc}
 \end{subfigure}
 \begin{subfigure}[b]{0.65\linewidth}
        \centering
		\input{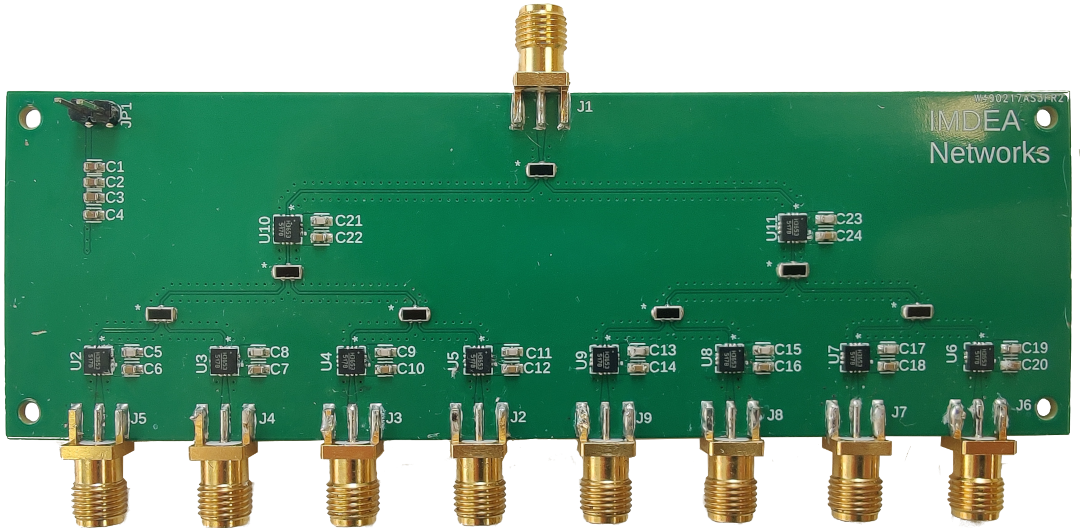}
        \caption{LO distribution.}
		\label{fig:lodist}
 \end{subfigure}
 \vspace{10pt}
 \begin{subfigure}[b]{0.28\linewidth}
        \centering
		\input{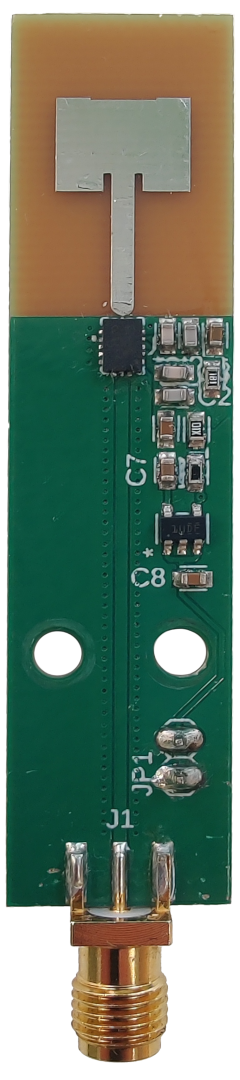}
        \caption{Active antenna.}
		\label{fig:ant}
 \end{subfigure}
 \vspace{10pt}
\caption{Hardware modules. Highlighting in black the HMC220B mixer, in red the GVA-123+ amplifiers biased by the TCBT-123+ bias tees, in blue the hfcw-9000+ and LFCN-113+ filters, in yellow the HMC3653LP3BE power amplifiers, in magenta the PS3216GT2 power splitters and in cyan the PMA-183PLN+ low noise amplifier.}
\label{fig:hardware}
\end{figure}

\begin{figure}[t]
    \centering
\begin{tikzpicture}

\begin{axis}[%
width=\fwidth,
height=\fheight,
scale only axis,
xmin=8,
xmax=12.5,
ymin=-15,
ymax=0,
xlabel near ticks,
ylabel near ticks,
ylabel={Gain [dB]},
xlabel={Frequency [GHz]}
]

\addplot [color=mycolor1, line width=2.0pt]
 table[row sep=crcr] {%
7	-40.0072173395276\\
7.12	-36.8816037147274\\
7.24	-34.2194908306019\\
7.36	-31.8244960657878\\
7.48	-29.5971807316124\\
7.6	-27.4395517695936\\
7.72	-25.2913201485238\\
7.84	-23.1222399604447\\
7.96	-20.8881747219741\\
8.08	-18.5670479969289\\
8.2	-16.1614808960776\\
8.32	-13.6433195817175\\
8.44	-11.0881409261372\\
8.56	-8.61036511023336\\
8.68	-6.41131271421576\\
8.8	-4.71658818711922\\
8.92	-3.68451722386095\\
9.04	-3.15428917351513\\
9.16	-2.86230951741846\\
9.28	-2.64104594968805\\
9.4	-2.50948471652069\\
9.52	-2.32708959481578\\
9.64	-2.17931969609773\\
9.76	-2.04423222898893\\
9.88	-1.93447919017345\\
10	-1.86290286336527\\
10.12	-1.84181311546294\\
10.24	-1.83405323222467\\
10.36	-1.87967148099826\\
10.48	-1.98136760026328\\
10.6	-2.12306458133639\\
10.72	-2.31198316124682\\
10.84	-2.56017007993721\\
10.96	-2.81178716766196\\
11.08	-3.05254543625152\\
11.2	-3.28200524345428\\
11.32	-3.45966798779852\\
11.44	-3.58273601886141\\
11.56	-3.73042948362445\\
11.68	-4.0205371184906\\
11.8	-4.60258070783661\\
11.92	-5.72073585026344\\
12.04	-7.46966566332698\\
12.16	-9.70062495013974\\
12.28	-12.2272874635108\\
12.4	-14.935872861482\\
12.52	-17.7170106092644\\
12.64	-20.6037881298727\\
12.76	-23.6168488288894\\
12.88	-26.8920497006443\\
13	-30.6020441351248\\
};

\end{axis}

\end{tikzpicture}%
    \caption{Bandpass filter response.}
    \label{fig:bpf_response}
 \end{figure}
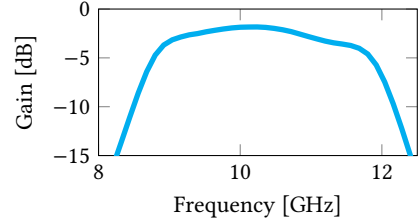

To perform the \gls{fr3} phase measurement, we designed and realized modular coherent hardware that allows us to up/down convert sub-$6$~GHz radios to the $9$~GHz-$12$~GHz band. The hardware setup, depicted in Fig.~\ref{fig:hardware}, is composed of four different types of \glspl{pcb}: 

\textbf{A Downconverter} \edit{ that converts a signal in the $9$~GHz-$12$~GHz range into a sub-$6$~GHz signal. In particular, first the received signal is amplified using a GVA-123+ amplifier \cite{gva123} (biased by a TCBT-123+ bias tee \cite{tcbt123}) to minimize the system's noise figure.} Subsequently, the signal is filtered by a \gls{bpf} stage  to remove out-of-band signals, particularly in the lower side band of the mixer. The filtering stage is composed of an HFCW-9000+ \gls{hpf} \cite{HFCW} and an LFCN-113+ \gls{lpf} \cite{LFCN}. The resulting filter response is depicted in Fig.~\ref{fig:bpf_response}, which shows that the usable frequency range is $9$~GHz to $12$~GHz. Note that there are pin-compatible filters with different cutoff frequencies, thus, this band can be extended. Then, the signal is further amplified to counteract the losses from the filters and the mixer. Finally, the signal is mixed with an externally generated \gls{lo} via an HMC220B mixer \cite{hmc220b} to generate the \gls{if} signal.

\textbf{An Upconverter} board, \edit{which takes a sub-$6$~GHz input \gls{if} signal, feeds it through the same chain as the downconverter in the opposite direction, and finally amplifies the $9$~GHz-$12$~GHz } signal with a HMC3653LP3BE power amplifier \cite{hmc3653} that has a \gls{p1db} of $15$~dBm at $10$~GHz, and is therefore capable of transmitting at $>10$~dBm without significant distortion.

\textbf{An LO distribution board} that ensures coherence between the \glspl{rx}. It utilizes \edit{PS3216GT2 resistive power splitter \cite{ps3216} to split} a single external \gls{lo} into $8$ channels that can be fed to as many downconverters or upconverters. The \gls{lo} \edit{is also amplified with HMC3653LP3BE amplifiers} between splits to maintain a \edit{suitable} power level for feeding the HMC220B mixer \cite{hmc220b}, which requires a typical \gls{lo} power of $10$~dBm, with a minimum of $7$~dBm. \edit{In particular, the signal is first split into two, then amplified, then further split twice to obtain the final $8$ channels, and finally amplified again.}

\textbf{An active antenna} \edit{module, which comprises a custom}  patch antenna and a PMA183-PLN+ low noise amplifier \cite{pma183} to overcome the cable losses and maintain a low system noise figure when using long cables for the distributed array. A passive version of the patch antenna is \edit{also available} for the \gls{tx} side. \edit{For this antenna, we measured a phase variation of $3.3^\circ$ RMS when rotating it around its phase center between $\pm 75^\circ$ from the boresight.}

All of the presented hardware is made freely available at \cite{frontends_link}. The \glspl{pcb} are designed for regular low-cost \gls{fr4} material, which despite not being intended for high frequency applications, has demonstrated sufficient performance for this experimental setup. The \glspl{pcb} can therefore be easily and cheaply manufactured through many online services, making this a very accessible experimental solution for the community. The components are also readily available from major distributors, and add up to a cost below $100$€ for the up and down converter, below $50$€ for the active antenna and below $400$€ for the \gls{lo} distribution board.

\section{Results} \label{sec:results}

\begin{figure}[t]
\centering
\input{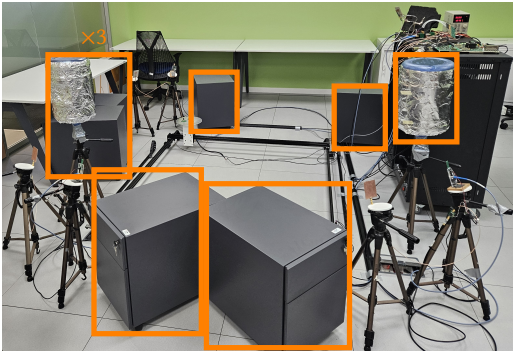}
\caption{Measurement setup with $8$ additional reflectors (highlighted in orange).}
\label{fig:physical_setup_MP}
\end{figure}

We perform an extensive characterization of the method by moving the \gls{tx} antenna in a $2$D square grid with a $5$~mm spacing between points, over the entire $1.2~\text{m} \times 1.2$~m area of the gantry,
\edit{(i.e., the \gls{tx} is placed at coordinates of the form $(k_1 \times 5\text{mm}, k_2 \times 5\text{mm},0)$, with $k_1$ and $k_2$ integers)}, 
for a total of ${\sim}58$k locations. The evaluation is repeated six times, resulting in ${\sim}350$k data points. The first and second runs are performed in the environment as depicted in Fig.~\ref{fig:physical_setup}. To ensure a diverse range of multipath propagation environments, for runs $3$ to $6$ we introduce $2$, $4$, $6$ and $8$ metallic objects, respectively, placed around the measurement area . The environment with $8$ metallic objects is depicted in Fig.~\ref{fig:physical_setup_MP}, with the reflectors fully covering each side of the measurement area (excluding the top side, where objects cannot be placed due to the moving \gls{tx} antenna cables). From the first run, we extract $N_c$ ($N_c=50$, unless specified otherwise) points uniformly spread over the measurement area for calibration.
%, which are used to evaluate the performance for all runs.
The results are organized as follows. In Sec.~\ref{sec:results_ref} we focus on the performance of the refinement method itself, to demonstrate that localization refinement is feasible with a sufficiently accurate initial estimate. Here, we use a synthetic initial estimate generated by adding Gaussian noise to the ground truth. In Sec.~\ref{sec:results_orientation} we discuss the impact of the antenna orientation, and demonstrate how accuracy is maintained despite the changes and imperfect knowledge of the orientation of the \gls{tx} antenna module. In Sec.~\ref{sec:results_tdoa} we show how appropriate tracking and sensor fusion can achieve a sufficiently accurate estimate to initialize the system \edit{and compare the performance with a \gls{tdoa}-based baseline.
}

\subsection{Refinement performance evaluation} \label{sec:results_ref}

\begin{figure}[t]
    \centering
    \input{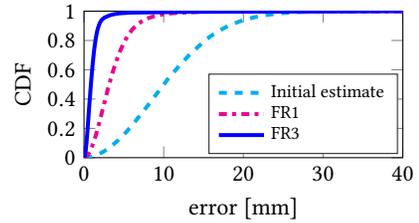}
    \caption{Accuracy \gls{cdf} at different stages of the refinement process with $N_c=50$ and $\epsilon=10$~mm.}
\label{fig:cdf_example}
\end{figure}

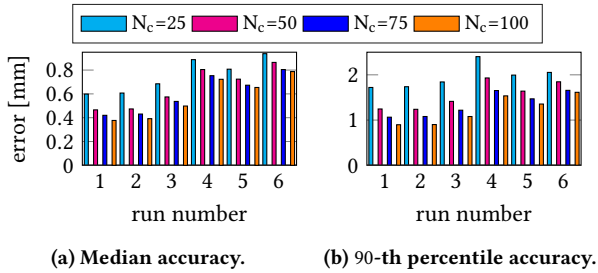
\begin{figure}[t]
    \centering
    \begin{subfigure}[t]{0.98\linewidth}
    \centering

%
\begin{tikzpicture}

\begin{axis}[%
width=0,
height=0,
scale only axis,
axis background/.style={fill=white},
legend style={at={(1.05,0.5)}, anchor= west, legend cell align=center, align=center, draw=white!15!black,  legend columns=4,font=\small}
]
\addplot[ybar, bar width=2, fill=mycolor1, draw=black, area legend] table[row sep=crcr] {%
0	0\\
0	0\\
};
\addplot[forget plot, color=white!15!black] table[row sep=crcr] {%
0	0\\
0	0\\
};
\addlegendentry{$\text{N}_\text{c}\text{=25}$}

\addplot[ybar, bar width=2, fill=mycolor2, draw=black, area legend] table[row sep=crcr] {%
0	0\\
0	0\\
};
\addplot[forget plot, color=white!15!black] table[row sep=crcr] {%
0	0\\
0	0\\
};
\addlegendentry{$\text{N}_\text{c}\text{=50}$}

\addplot[ybar, bar width=2, fill=mycolor3, draw=black, area legend] table[row sep=crcr] {%
0	0\\
0	0\\
};
\addplot[forget plot, color=white!15!black] table[row sep=crcr] {%
0	0\\
0	0\\
};
\addlegendentry{$\text{N}_\text{c}\text{=75}$}

\addplot[ybar, bar width=2, fill=mycolor4, draw=black, area legend] table[row sep=crcr] {%
0	0\\
0	0\\
};
\addplot[forget plot, color=white!15!black] table[row sep=crcr] {%
0	0\\
0	0\\
};
\addlegendentry{$\text{N}_\text{c}\text{=100}$}

\end{axis}

\end{tikzpicture}%
    \end{subfigure}
    \begin{subfigure}[t]{0.48\linewidth}
     \centering

%
\begin{tikzpicture}

\begin{axis}[%
width=\cbwith,
height=\cbheight,
scale only axis,
bar shift auto,
xmin=0.5,
xmax=6.5,
xtick={1, 2, 3, 4, 5, 6},
xlabel style={font=\color{white!15!black}},
xlabel={run number},
xlabel near ticks,
xtick style={draw=none},
ymin=0,
ymax=0.95,
ylabel style={font=\color{white!15!black}},
ylabel={error [mm]},
ylabel near ticks,
axis background/.style={fill=white},
legend style={at={(1.05,0.5)}, anchor= west, legend cell align=left, align=left, draw=white!15!black}
]
\addplot[ybar, bar width=1.5, fill=mycolor1, draw=black, area legend, ] table[row sep=crcr] {%
1	0.59837733998937\\
2	0.607120780972036\\
3	0.684710495135195\\
4	0.887614527723282\\
5	0.807473552901759\\
6	0.937811909298805\\
};
\addplot[forget plot, color=white!15!black] table[row sep=crcr] {%
0.509090909090909	0\\
6.49090909090909	0\\
};
%\addlegendentry{$\text{N}_\text{c}\text{=25}$}

\addplot[ybar, bar width=1.5, fill=mycolor2, draw=black, area legend] table[row sep=crcr] {%
1	0.465755218131996\\
2	0.473649938071552\\
3	0.574413546914212\\
4	0.804490268244889\\
5	0.723829371410067\\
6	0.864429707365409\\
};
\addplot[forget plot, color=white!15!black] table[row sep=crcr] {%
0.509090909090909	0\\
6.49090909090909	0\\
};
%\addlegendentry{$\text{N}_\text{c}\text{=50}$}

\addplot[ybar, bar width=1.5, fill=mycolor3, draw=black, area legend] table[row sep=crcr] {%
1	0.420122025806894\\
2	0.430716155460507\\
3	0.536766601345984\\
4	0.752852714178376\\
5	0.673010478255031\\
6	0.804497710135452\\
};
\addplot[forget plot, color=white!15!black] table[row sep=crcr] {%
0.509090909090909	0\\
6.49090909090909	0\\
};
%\addlegendentry{$\text{N}_\text{c}\text{=75}$}

\addplot[ybar, bar width=1.5, fill=mycolor4, draw=black, area legend] table[row sep=crcr] {%
1	0.377037329576532\\
2	0.391975905469394\\
3	0.498027255762467\\
4	0.72260382926429\\
5	0.653983294355731\\
6	0.787433917083948\\
};

%\addlegendentry{$\text{N}_\text{c}\text{=100}$}

\end{axis}

\end{tikzpicture}%
 %updated and y-axis changed
            \caption{Median accuracy.}
            \label{fig:median_err_cal}
    \end{subfigure}
    \begin{subfigure}[t]{0.48\linewidth}
     \centering

%
\begin{tikzpicture}

\begin{axis}[%
width=\cbwith,
height=\cbheight,
scale only axis,
bar shift auto,
xmin=0.5,
xmax=6.5,
xtick={1, 2, 3, 4, 5, 6},
xlabel style={font=\color{white!15!black}},
xlabel={run number},
xlabel near ticks,
xtick style={draw=none},
ymin=0,
ymax=2.5,
% ylabel style={align=center,font=\color{white!15!black}},
% ylabel={error [mm]},
% ylabel near ticks,
axis background/.style={fill=white},
legend style={at={(1.05,0.5)}, anchor= west, legend cell align=left, align=left, draw=white!15!black}
]
\addplot[ybar, bar width=1.5, fill=mycolor1, draw=black, area legend] table[row sep=crcr] {%
1	1.72089974847848\\
2	1.73723539634281\\
3	1.84283045239726\\
4	2.40187404574298\\
5	1.99420792394476\\
6	2.05361089379985\\
};
\addplot[forget plot, color=white!15!black] table[row sep=crcr] {%
0.509090909090909	0\\
6.49090909090909	0\\
};
%\addlegendentry{$\text{N}_\text{c}\text{=25}$}

\addplot[ybar, bar width=1.5, fill=mycolor2, draw=black, area legend] table[row sep=crcr] {%
1	1.24555002417842\\
2	1.23827221389775\\
3	1.41343263637234\\
4	1.93098626481688\\
5	1.63977848154901\\
6	1.84588945993805\\
};
\addplot[forget plot, color=white!15!black] table[row sep=crcr] {%
0.509090909090909	0\\
6.49090909090909	0\\
};
%\addlegendentry{$\text{N}_\text{c}\text{=50}$}

\addplot[ybar, bar width=1.5, fill=mycolor3, draw=black, area legend] table[row sep=crcr] {%
1	1.06407549824615\\
2	1.07914979924988\\
3	1.21855337277124\\
4	1.65280030477288\\
5	1.46944746353511\\
6	1.6573466641433\\
};
\addplot[forget plot, color=white!15!black] table[row sep=crcr] {%
0.509090909090909	0\\
6.49090909090909	0\\
};
%\addlegendentry{$\text{N}_\text{c}\text{=75}$}

\addplot[ybar, bar width=1.5, fill=mycolor4, draw=black, area legend] table[row sep=crcr] {%
1	0.896957729635653\\
2	0.901177399415346\\
3	1.07880512691209\\
4	1.53522635790397\\
5	1.35464927679672\\
6	1.61380923488189\\
};
\addplot[forget plot, color=white!15!black] table[row sep=crcr] {%
0.509090909090909	0\\
6.49090909090909	0\\
};
%\addlegendentry{$\text{N}_\text{c}\text{=100}$}

\end{axis}

\end{tikzpicture}%
 %updated
        \caption{$90$-th percentile accuracy.}
        \label{fig:90th_err_cal}
    \end{subfigure}
    \vspace{10pt}
    \caption{Accuracy as a function of the number of calibration points.}
    \label{fig:cdf_vs_cal_synth}
\end{figure}

To evaluate the performance of the refinement method independently from the initial location estimate, in this section we generate the initial estimate $\hat{x}_0$ by adding an independent white Gaussian noise to the ground truth. We quantify the initial estimate error by its median, that we call $\epsilon$, which unless specified otherwise is set to $\epsilon=10$~mm to ensure the performance is limited by the refinement algorithm. The impact of the initial localization accuracy is discussed later and how the method can be realistically applied is shown in Sec.~\ref{sec:results_tdoa}. Fig.~\ref{fig:cdf_example} shows the aggregate \gls{cdf} of the achieved accuracy at each stage of the refinement process. The distribution of the error of the synthetic initial estimate shows a median error of $1$~cm, as designed. After the first refinement stage using the \gls{fr1} band, we obtain a median error of approximately $2.5$~mm, and after the final refinement with the \gls{fr3} band the median error drops to around $600 \: \mu$m.  Note that here, as well as in all other results, the ground truth is taken as the position we input to the gantry. Given the specified accuracy of the gantry is $200 \: \mu$m, part of the error might be due to the ground truth limitations.
In Fig.~\ref{fig:cdf_vs_cal_synth} we show the median accuracy of the final refinement stage for all multipath environments and for different numbers of calibration points. Here we can clearly see that both the multipath and the number of calibration points have a limited impact, with a $ < 2.5 \times$ performance degradation from the best to the worst case. Overall, the median accuracy ranges between ${\sim}370 \: \mu$m and ${\sim}940 \: \mu$m and the $90$-th percentile between ${\sim}0.9$~mm and ${\sim}2.4$~mm, thus confirming that the method can achieve one-shot millimetric accuracy if initialized with a sufficiently accurate estimate. We recall that we use the same calibration function, measured in the absence of additional reflectors, in all scenarios. This highlights how the calibration function is largely independent from the environmental changes. In other words, the calibration function is indeed mainly capturing imperfections in the knowledge of the antenna position and center of phase variations, rather than phase variations that are caused by multipath.

In addition, we found that the phase measurement accuracy in our setup is much lower than the method's requirements. \edit{In particular, we measured the phase for a fixed location $2000$ times over a time span of $30$ minutes, and observed an RMS phase variation $\leq 1.5^\circ$ for \gls{fr3} and $\leq 0.4^\circ$ for \gls{fr1}. Furthermore,} we observed ${\sim}1.2^\circ$ and ${\sim}1.5^\circ$ \gls{rms} difference edit{over all locations} between run $1$ and run $2$ for \gls{fr1} and \gls{fr3} respectively. ${\sim}14.9^\circ$  and ${\sim}16.5^\circ$ \gls{rms} difference for each band, comparing runs $1$ and $4$. \edit{In view of this,} we investigate the robustness of the method to higher phase error by adding an \emph{artificial} \gls{iid} Gaussian noise with standard deviation $\sigma_p$ to the phase measurements (with identical statistics for both bands). The resulting accuracies are shown in Fig.~\ref{fig:bars_phase_err}, which shows how even with an \textit{additional} phase error with $\sigma_p = 10^\circ$ (which adding the intrinsic error corresponds to a total error between $13^\circ$ and $21^\circ$) we can maintain median accuracy and a $90$-th percentile accuracy below $1.1$~mm and $2.3$~mm, respectively.

\begin{figure}[t]
    \centering
     \begin{subfigure}[t]{0.98\linewidth}
    \centering
        \input{figures/results/cdf_vs_phaseerr} %updated
        \caption{\gls{cdf} at each refinement step.}
        \vspace{13pt}
         \label{fig:cdf_vs_phaseerr}
    \end{subfigure}
    \begin{subfigure}[t]{0.98\linewidth}
    \centering

%
\begin{tikzpicture}

\begin{axis}[%
width=0,
height=0,
scale only axis,
axis background/.style={fill=white},
legend style={at={(1.05,0.5)}, anchor= west, legend cell align=left, align=left, draw=white!15!black,  legend columns=4,font=\small}
]
\addplot[ybar, bar width=2, fill=mycolor1, draw=black, area legend] table[row sep=crcr] {%
0	0\\
0	0\\
};
\addplot[forget plot, color=white!15!black] table[row sep=crcr] {%
0	0\\
0	0\\
};
\addlegendentry{$\sigma_p=5^\circ$}

\addplot[ybar, bar width=2, fill=mycolor2, draw=black, area legend] table[row sep=crcr] {%
0	0\\
0	0\\
};
\addplot[forget plot, color=white!15!black] table[row sep=crcr] {%
0	0\\
0	0\\
};
\addlegendentry{$\sigma_p=10^\circ$}

\addplot[ybar, bar width=2, fill=mycolor3, draw=black, area legend] table[row sep=crcr] {%
0	0\\
0	0\\
};
\addplot[forget plot, color=white!15!black] table[row sep=crcr] {%
0	0\\
0	0\\
};
\addlegendentry{$\sigma_p=20^\circ$}

\end{axis}

\end{tikzpicture}%
    \end{subfigure}
    \begin{subfigure}[t]{0.48\linewidth}
     \centering

%
\begin{tikzpicture}

\begin{axis}[%
width=\cbwith,
height=\cbheight,
scale only axis,
bar shift auto,
xmin=0.5,
xmax=6.5,
xtick={1, 2, 3, 4, 5, 6},
xlabel style={font=\color{white!15!black}},
xlabel={run number},
xlabel near ticks,
xtick style={draw=none},
ymin=0,
ymax=1.65,
ylabel style={font=\color{white!15!black}},
ylabel={error [mm]},
ylabel near ticks,
axis background/.style={fill=white},
legend style={at={(1.05,0.5)}, anchor= west, legend cell align=left, align=left, draw=white!15!black}
]
\addplot[ybar, bar width=1.5, fill=mycolor1, draw=black, area legend] table[row sep=crcr] {%
1	0.597485904053099\\
2	0.602253952002323\\
3	0.680172333408036\\
4   0.875357674130431\\
5	0.804883632603978\\
6	0.931713404980409\\
};
\addplot[forget plot, color=white!15!black] table[row sep=crcr] {%
0.509090909090909	0\\
6.49090909090909	0\\
};
%\addlegendentry{$\sigma_p=5^\circ$}

\addplot[ybar, bar width=1.5, fill=mycolor2, draw=black, area legend] table[row sep=crcr] {%
1	0.840309861409684\\
2	0.846919925853033\\
3	0.90050443686047\\
4	1.05370854463433\\
5	0.999970833194352\\
6	1.09757033150448\\
};
\addplot[forget plot, color=white!15!black] table[row sep=crcr] {%
0.509090909090909	0\\
6.49090909090909	0\\
};
%\addlegendentry{$\sigma_p=10^\circ$}

\addplot[ybar, bar width=1.5, fill=mycolor3, draw=black, area legend] table[row sep=crcr] {%
1	1.40751929498803\\
2	1.40319877643851\\
3	1.45555148616614\\
4	1.57954616864068\\
5	1.51550703113844\\
6	1.59384518040697\\
};
\addplot[forget plot, color=white!15!black] table[row sep=crcr] {%
0.509090909090909	0\\
6.49090909090909	0\\
};
%\addlegendentry{$\sigma_p=20^\circ$}

\end{axis}
\end{tikzpicture}%
%updated
            \caption{Median accuracy.}
            \label{fig:median_err_phase}
    \end{subfigure}
    \begin{subfigure}[t]{0.48\linewidth}
     \centering

%
\begin{tikzpicture}

\begin{axis}[%
width=\cbwith,
height=\cbheight,
scale only axis,
bar shift auto,
xmin=0.5,
xmax=6.5,
xtick={1, 2, 3, 4, 5, 6},
xlabel style={font=\color{white!15!black}},
xlabel={run number},
xlabel near ticks,
xtick style={draw=none},
ymin=0,
ymax=3.5,
% ylabel style={font=\color{white!15!black}},
% ylabel={error [mm]},
% ylabel near ticks,
axis background/.style={fill=white},
legend style={at={(1.05,0.5)}, anchor= west, legend cell align=left, align=left, draw=white!15!black}
]
\addplot[ybar, bar width=1.5, fill=mycolor1, draw=black, area legend] table[row sep=crcr] {%
1	1.36149140680026\\
2	1.35997691445279\\
3	1.51188601803804\\
4	2.0216765863548\\
5	1.73378287619883\\
6	1.93677117956702\\
};
\addplot[forget plot, color=white!15!black] table[row sep=crcr] {%
0.509090909090909	0\\
6.49090909090909	0\\
};
%\addlegendentry{$\sigma_p=5^\circ$}

\addplot[ybar, bar width=1.5, fill=mycolor2, draw=black, area legend] table[row sep=crcr] {%
1	1.69378986577369\\
2	1.69174906484821\\
3	1.8222067310934\\
4	2.30774996791628\\
5	2.02501676438616\\
6	2.20798605207736\\
};
\addplot[forget plot, color=white!15!black] table[row sep=crcr] {%
0.509090909090909	0\\
6.49090909090909	0\\
};
%\addlegendentry{$\sigma_p=10^\circ$}

\addplot[ybar, bar width=1.5, fill=mycolor3, draw=black, area legend] table[row sep=crcr] {%
1	2.68410880193849\\
2	2.67092372635197\\
3	2.80976903444078\\
4	3.37457753566945\\
5	2.99026136301767\\
6	3.15621419621743\\
};
\addplot[forget plot, color=white!15!black] table[row sep=crcr] {%
0.509090909090909	0\\
6.49090909090909	0\\
};
%\addlegendentry{$\sigma_p=20^\circ$}

\end{axis}

\end{tikzpicture}%
%updated
        \caption{$90$-th percentile accuracy.}
        \label{fig:90th_err_phase}
    \end{subfigure}
    \vspace{10pt}
    \caption{Accuracy as a function of the \emph{additional} synthetic phase error $\sigma_p$ (beyond the $1.5^\circ$ to $16.5^\circ$ intrinsic to the measurement).}
    \label{fig:bars_phase_err}
\end{figure}
Fig.~\ref{fig:cdf_vs_phaseerr} shows the \gls{cdf} of the accuracy achieved \edit{when} \emph{additional} phase error is introduced. Here we can see that up to $\sigma_p = 20^\circ$ the degradation is mainly due to the phase error at the $10.25$~GHz band, and the $3.25$~GHz is still accurate enough to select the correct peak at $10.25$~GHz. To push the refinement to its failure point, we needed to add a considerable \textit{additional} phase error of $\sigma_p = 30^\circ$ (for a total upwards of $40^\circ$). Doing so we start to observe a large enough degradation in the $3.25$~GHz band to misidentify the likelihood peak at \gls{fr3}, which is highlighted by the long tail of the \gls{cdf}.\footnote{Note, we do not show the $\sigma_p = 30^\circ$ case in Fig.~\ref{fig:bars_phase_err} as the long tail in the \gls{cdf} would make the $90$-th percentile plot difficult to read.} As discussed in \ref{sec:phase_meas}, such error is highly unlikely even in extreme multipath. 

\begin{figure}[t]
     \centering
     \input{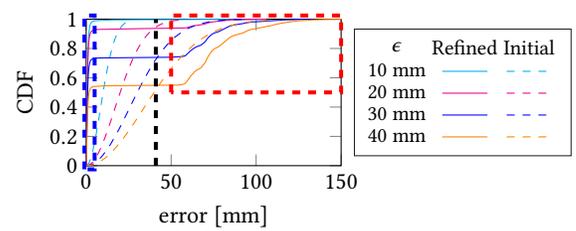}%updated
     \caption{\gls{cdf} of localization accuracy for varying initial estimation errors.}
     \label{fig:cdf_vs_initerr}
\end{figure}

Finally, Fig.~\ref{fig:cdf_vs_initerr} illustrates the impact of the median initial localization error $\epsilon$ on the final localization accuracy. The \gls{cdf} reveals a distinct step-like behavior, which highlights two key possibilities. Highlighted in blue are the cases where the initial estimate is accurate enough to identify the correct peak of the \gls{fr1} likelihood function, leading to final errors in the millimeter range, and highlighted in red are the cases where instead the peak identification fails, resulting in significantly larger errors. Interestingly, the transition point between these two regimes aligns consistently with an initial estimation error of approximately $41$~mm (marked with the dashed black line in the figure), with minor fluctuations. This observation allows us to impose a statistical constraint on the accuracy requirements of the initial localization system. Moreover, this step-like behavior in the \gls{cdf}, which could at first glance look detrimental to the method's robustness, is in actuality advantageous, as it enables straightforward outlier detection. 

\subsection{Impact of orientation} \label{sec:results_orientation}

As discussed in Sec.~\ref{sec:likelihood_calc}, evaluating the likelihood function requires knowledge of the orientation of the \gls{tx} antenna module. To assess the sensitivity to inexact knowledge of such orientation, we introduce an artificial Gaussian \gls{iid} noise with standard deviation $\sigma_o$ to the initial estimate of the orientation. Here we see how the results with larger angular error, presented in Fig.~\ref{fig:orientation}, exhibit a similar behavior to that with an incorrect initial position estimate, resulting in a sharp transition in the \gls{cdf}. However, while the observed behavior is similar, the underlying mechanism differs. When the initial estimate is incorrect, the correct likelihood peak at \gls{fr1} is misidentified, leading to a final estimate error on the order of the distance between two likelihood peaks at $3.25$~GHz (over $6$~cm, as shown in Fig.~\ref{fig:cdf_vs_initerr}). In contrast, when the orientation is incorrect, the initial refinement typically succeeds, as the ambiguity at \gls{fr1} is forgiving. However, the first refined estimate $\hat{x}_1$ suffers from an error that is due to the incorrect direction of the displacement vector $\Delta_1$. This can then cause the second refinement step to select an incorrect peak. As a result, in this case the final localization error when the peak is misidentified corresponds to the distance between two likelihood peaks at the highest frequency (${\sim}2.5$~cm, as seen in Fig.~\ref{fig:cdf_vs_initerr}).  Further analysis of Fig.~\ref{fig:orientation} reveals that the transition in the \gls{cdf} occurs at probabilities $p = 0.96, 0.86, 0.75$ for $\sigma_o = 4^\circ, 6^\circ, 8^\circ$, respectively. Using the inverse Gaussian \gls{cdf}, these probabilities correspond to errors of approximately $2.05$, $1.47$, and $1.15$ standard deviations. Multiplying these values by the respective $\sigma_o$ yields angular errors of $8.2^\circ$, $8.82^\circ$, and $9.02^\circ$. This suggests that statistically the maximum tolerable orientation error is approximately ${\sim}8.5^\circ$. 
In Sec.~\ref{sec:results_tdoa}, we show that with calibration, a commercial magnetometer can provide orientation with an accuracy of ~$0.86^\circ$, widely below the computed tolerance.
\begin{figure*}[t]
    \centering
    \begin{minipage}{0.28\linewidth}
        \centering
		\input{figures/results/cdf_vs_angle_err}
        \caption{Localization error \gls{cdf} as a function of the standard deviation of the orientation error.}
		\label{fig:orientation}
    \end{minipage}
    \hfill
    \begin{minipage}{0.34\linewidth}
        \centering
        \input{figures/results/rotator}%Updated 
        \caption{Estimated location for different \gls{tx} orientations.}
        \label{fig:rotator_test}
    \label{fig:rotator_test}
    \end{minipage}
        \hfill
    \begin{minipage}{0.34\linewidth}
        \centering
		\input{figures/results/mobility_trajectory}
        \caption{Trajectory of the \gls{tx} under real mobility.}
		\label{fig:mobility}
    \end{minipage}
\end{figure*}
Moreover, changes in orientation of the \gls{tx} antenna module could cause degradation of the performance due to variations of the observed center of phase \cite{phase_center_1, phase_center_2}. As all measurements presented so far are done with a fixed orientation of the \gls{tx} module, additional measurements are required to evaluate the performance under orientation changes. For this evaluation, we mount the module on a FLIR PTU-D47 \cite{PTUD47} rotating platform to adjust the \gls{tx} module orientation with a resolution of ${\sim}0.05^\circ$ over a range of $\pm 150^\circ$ in the azimuthal plane. The antenna is rotated through the full range in steps of ${\sim}1^\circ$, and the proposed method is used to estimate its position. Since the antenna module is rotated about its center, and has the $10.25$~GHz antenna located ${\sim}50$~mm from the center, the expected trajectory of the estimated locations should form a circle with a ${\sim}50$~mm radius. Note that both the placement of the rotating platform and the assembly of the antenna module are done manually. Thus, the best measurement we have of both the module initial position and the distance between the center and the antenna are the one provided by the interferometry itself. As such, the reported error is calculated as the distance between the measured point and the best fitting circle. The measured locations obtained are shown in Fig.~\ref{fig:rotator_test}, alongside the best fit circle depicted in black. The results closely follow the expected circular trajectory, confirming that the proposed method remains effective. The error \gls{cdf} shows that millimeter accuracy is maintained, with a median error of $900 \: \mu$m.

\subsection{Robustness of the method}
\edit{In this section, we assess the resilience of the method to \gls{tx} mobility, dynamic environments, \gls{rx} antenna placement and temperature changes by designing a new experiment. In this experiment, the \gls{tx} is not stopped at each point, but rather moves at a constant velocity of $82.5\frac{\text{mm}}{\text{s}}$ (limited by the speed of the gantry) over a pre-defined trajectory, and we collect the channel estimate every $20$~ms. Further, to demonstrate that the method works with different $3$D configurations of \gls{rx} antenna position, we move the \gls{rx} antennas to different heights, with the $10.25$~GHz antennas ranging from $40$~cm to $1$~m above the \gls{tx} movement plane, and the $3.25$~GHz antennas between $30$~cm and $82$~cm. To demonstrate the robustness of the method to realistic dynamic environments, during this measurement, we had two volunteers performing their daily lab activities close to the experiment while the data was collected (at $1$~m and $2$~m from the closest RX antenna respectively). The ground truth and estimated trajectory in these conditions are depicted in Fig.~\ref{fig:mobility}. Note that a slight deviation in the synchronization between the motors and the acquisition might lead to a significant error in the ground truth (e.g., with a synchronization error of only $10$~ms, given the speed of the \gls{tx}, the ground truth would be shifted by $0.82$~mm, which is comparable to the system's accuracy). 
For this reason, we mathematically generate the ground truth trajectory with a granularity of $1$~$\mu$m in both the $x$ and $y$ axes, we associate each estimated point with the closest point of the ground truth trajectory, and we compute the RMSE, obtaining a value of $987$~$\mu$m.
Further, to evaluate how ambient temperature impacts the system, we re-run the the same trajectory at night time temperature ($\sim 18^\circ$C, while the calibration and first run where done at $\sim 22^\circ$C). In this condition, we obtain a $1.16$~mm RMSE, showing that day-night cycle temperature changes cause minimal degradation. }

\subsection{Use case evaluation} \label{sec:results_tdoa}
The proposed location refinement method relies on the availability of sufficiently accurate initial position estimates to refine them to millimeter-level accuracy. 

In this section, we show that for the example use case of localizing a robot along a trajectory, it is possible to obtain accurate enough initial position estimates through the combination of \gls{sota} methods.
Moreover, we use these initial position estimates as a baseline for comparison with our proposed method, as we consider it a good representation of the achievable localization performance with current \gls{sota} solutions
(the bounds in \cite{bounds_1} demonstrate that no baseband \gls{5g} localization system can achieve significantly better performance).

\begin{figure}[t]
\centering
	\input{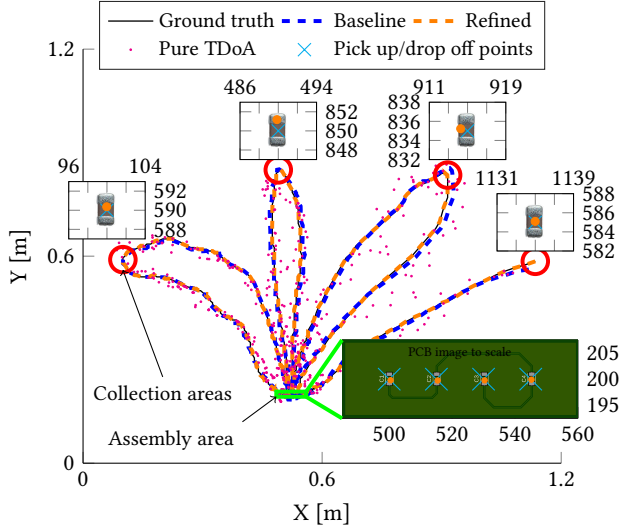}
        \caption{Example trajectory at different stages of the processing pipeline.}
	\label{fig:trajectory1}
\end{figure}

\begin{figure}[t]
\centering
		\input{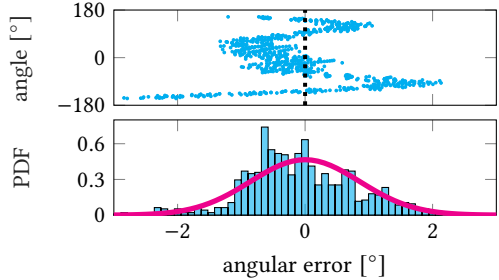}
        \caption{Measured HMC5883L magnetometer error.}
		\label{fig:rotator_magnetometer}
\end{figure}

\smallsection{Evaluation setup}
We consider the use case of a robot tasked with collecting some small \gls{smd} electronic components from different positions within the working area and placing them at precise target locations on a \gls{pcb}. The robot is equipped with a wheel encoder to measure the robot's velocity and a magnetometer to measure the robot's orientation. 
The robot trajectory is emulated by taking a sequence of points in the grid measurements presented in \secref{sec:platform}, and the encoder and magnetometer data are synthesized based on the characteristics of real sensors.
 For the velocity sensor's noise model we consider the quantization error introduced by the finite encoder steps, computed based on the specifications of a commercial (${\sim}100$~€) DLS40E-S3AV01024 wheel encoder \cite{wheel-encoder}. Assuming a zero-mean uniform distribution of the error, its standard deviation can be computed as $\sigma_{\Tilde{v}}=2\pi r/ (\sqrt{12} \cdot PPR \cdot T)\approx1$~mm/s, where $r=100$~mm is the robot wheel radius, $PPR=1024$ is the number of pulses per revolution offered by the encoder, and $T=175$~ms is the velocity estimation period. Note that this result is independent of the velocity.
Similarly, we \edit{emulate a realistic orientation sensor by experimentally evaluating the performance of a HMC5883L magnetometer (${\sim}10$~€) \cite{HMC5883L}, to then generate an equivalent orientation error}. 
For this evaluation, we mounted the magnetometer on the same rotating platform used in Sec.~\ref{sec:results_orientation}. We first run a full sweep to calibrate the magnetometer. The calibration is done by fitting the difference between the measured and ground truth angle with a $3$-coefficient Fourier model \cite{FitFourierModels}. We then sweep the angle three times to evaluate the accuracy. Fig.~\ref{fig:rotator_magnetometer} shows the magnetometer error as a function of the ground truth angle on the top and the histogram of the error on the bottom, along with the fitted Gaussian with a standard deviation of $\sigma_{\Tilde{\gamma}}\approx 0.86^\circ$ that we use for the experimental orientation readings. 

\smallsection{Baseline calculation}
Given the scenario described above, we implement a robot-tracking algorithm that fuses all the available sensor information, for enhanced robot positioning. 

We start from a standard \gls{tdoa} algorithm \cite{tdoa} operating in the \gls{fr3} band with the available $400$~MHz bandwidth.
As the \gls{tdoa} estimates can be affected by systematic impairments, we perform a calibration on the measured \gls{tdoa} using a strategy similar to the one used for the phase estimation. In particular, we compute the difference between the modeled and measured differential time of arrival at $N_c=100$ calibration points, and perform a \gls{loess} fit to obtain the calibration function, which is then included in the \gls{tdoa} model to compensate for systematic errors. 

Then, to fuse the information from the encoder, the magnetometer and the \gls{tdoa} estimates we adopt a \gls{kf}. We model our system as
$\vv{x}_m = \vv{F} \vv{x}_{m-1} + \vv{w}_m$ and 
$\vv{y}_m = \vv{H} \vv{x} + \vv{v}_m$,
where $m$ represents the discrete-time instant, $\vv{x}_m=[x_m, y_m, \dot{x}_m, \dot{y}_m]^T$ is the system state, composed of the robot's position and velocity components, respectively,  $\vv{y}_m$ is the measurements vector, $\vv{F}$ is the constant-velocity state-transition model \cite{wagner2017radar}, $\vv{H}=\vv{I}_4$ is the observation model, with $\vv{I}_N$ the N-dimensional identity matrix, and $\vv{w}_m$ and $\vv{v}_m$ are the white, zero-mean, uncorrelated process and observation noises, respectively. To avoid introducing nonlinearities in the observation model, we substitute the velocity $\Tilde{v}_m$ from the wheel encoder and the orientation $\Tilde{\gamma}_m$ from the magnetometer with \emph{pseudo-measurements} \cite{wagner2017radar} that preserve the linearity of the system, providing $\vv{y}_m = [\Tilde{x}_m, \Tilde{y}_m, \Tilde{v}\cos({\Tilde{\gamma}_m}), \Tilde{v}\sin({\Tilde{\gamma}_m})]^T$ and use a covariance of the process and observation noises of $\vv{Q} = \sigma_a \, \vv{I}_4$ and $\vv{R} = [\sigma_x, \sigma_y, \sigma_{\dot{x}}, \sigma_{\dot{y}}] \, \vv{I}_4$, respectively.
Empirically, we set $\sigma_x=\sigma_y=300$~mm, $\sigma_{\dot{x}}=\sigma_{\dot{y}}=5$~m/s, and $\sigma_a=50~\rm m/s^2$.

\begin{figure*}[t]
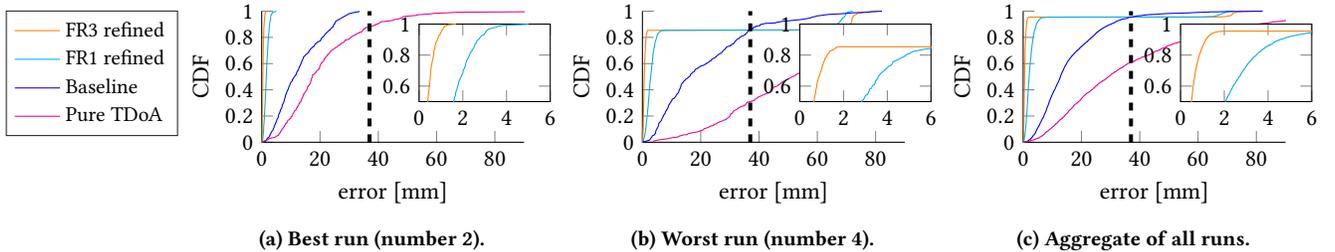

    \centering
    \begin{subfigure}[t]{0.13\textwidth}
    \centering

%
\begin{tikzpicture}

\begin{axis}[%
width=0,
height=0,
scale only axis,
axis background/.style={fill=white},
legend style={at={(0,1cm)}, anchor=south, legend cell align=left, align=center, draw=white!15!black,font=\small,
legend columns=1}
]
\addplot[color=mycolor4] table[row sep=crcr] {%
0	0\\
0	0\\
};
\addlegendentry{\gls{fr3} refined}

\addplot[color=mycolor1] table[row sep=crcr] {%
0	0\\
0	0\\
};
\addlegendentry{\gls{fr1} refined}

\addplot[color=mycolor3] table[row sep=crcr] {%
0	0\\
0	0\\
};
\addlegendentry{Baseline}

\addplot[color=mycolor2] table[row sep=crcr] {%
0	0\\
0	0\\
};
\addlegendentry{Pure \gls{tdoa}}

\end{axis}

% \draw[white] (0,0) ++ (0, -1.5cm); %trick to put the legend higher

\end{tikzpicture}%
        % \caption{\mc{Adjust legend.}}
    \end{subfigure}    
    \begin{subfigure}[t]{0.28\textwidth}
    \centering
        \input{figures/results/system_level/tdoa_5g_matrix_2}
        \caption{Best run (number 2).}
        \label{fig:run1_sys}
    \end{subfigure} 
    \begin{subfigure}[t]{0.28\textwidth}
     \centering
        \input{figures/results/system_level/tdoa_5g_matrix_4}
            \caption{Worst run (number 4).}
            \label{fig:run6_sys}
    \end{subfigure}
    \begin{subfigure}[t]{0.28\textwidth}
     \centering
        \input{figures/results/system_level/tdoa_5g_matrix_aggregate}
        \caption{Aggregate of all runs.}
        \label{fig:aggregate_sys}
    \end{subfigure}
    \vspace{10pt}
    \caption{Localization error \gls{cdf} along the trajectory.}
    \label{fig:localization_cdf_sys}
\end{figure*}

\smallsection{Results}
As previously explained, we use the filtered trajectory both as a baseline for comparison, as it represents the accuracy practically achievable with \gls{sota} algorithms, and as to obtain the initial estimates for our position refinement method.

The trajectory followed by the robot is depicted in \fig{fig:trajectory1} for different stages of the processing pipeline. The trajectory loops represent movements to collect the desired components, and the pick up and drop off areas are marked in red and green, respectively.
 
The bottom right of the figure shows a magnified view of the drop points superimposed on a to-scale view of an example \gls{pcb} that needs to be assembled, where we can clearly see that the accuracy of the estimate is sufficient for correctly installing the components. Similarly, on the top we can see a to-scale view of the 1206 \gls{smd} components being picked up, with the estimated location superimposed. Here, we also see that the pick up locations are sufficiently accurate to correctly collect the component.  

\fig{fig:localization_cdf_sys} shows the \gls{cdf} of the localization error along the whole trajectory using data from different runs. Specifically, we present the results of the run performing the best (run 2) and the worst (run 4), along with the aggregate of all runs.
In the best scenario, the median and $90$-th percentile are approximately $400~\mu \rm m$ and $840~\mu \rm m$, respectively. 
In the worst scenario the median error remains as low as $645~\mu \rm m$, while the $90$-th percentile error becomes high with a value of $71$~mm, because the probability at which the \gls{cdf} step happens (see discussion in \secref{sec:results_ref}) is $0.85 < 0.9$.
Further, the results of this example confirm the previously observed step-like behavior of the \gls{cdf}, with a consistent transition point corresponding to a specific initial estimation error, represented by the dashed black line in \fig{fig:localization_cdf_sys}. However, in this case the transition point drops down to $37$~mm, compared to the $41$~mm observed previously. Finally, the aggregate error shows a median of around $500~\mu \rm m$ and a $90$-th percentile of $1.16$~mm. From the same plot, we notice that our refinement method in FR3 remarkably shows a localization error $25$ times lower than the baseline in both median and $90$-th percentile.

Overall, these results show that the method maintains high accuracy even with realistic initial estimates, with a remarkable $25\times$ improvement over the baseline localization method, and prove that its accuracy is sufficient to perform high precision manufacturing tasks such as \gls{smd} assembly.

\section{Related Work}

To achieve high accuracy localization and sensing, a common solution is to stitch multiple bands together to generate an equivalent wideband channel \cite{wifi_band_stiutching, band_stitching_2}. These methods, however, do not exploit carrier phase information, and therefore suffer from the coupling between bandwidth and accuracy. In contrast, in our method the final accuracy is largely uncorrelated with the bandwidth (as long as the bandwidth is sufficient to discriminate the \gls{los}).
Previous works exploiting carrier phase information exists, but they suffer from limitations related to the difficulties in resolving the integer ambiguity.
Some methods resolve the ambiguity by tracking and unwrapping the phase over target movements \cite{tracking_1, tracking_2, tracking_3, tracking_4, tracking_5, mmvib}. As a result, those methods are only capable of precise \textit{relative} measurements, while our method is capable of providing accurate absolute localization.
Other methods rely on the consistency among multiple antennas \cite{phase_only_localization}, but this requires \gls{los} with a large number of antennas, which is unlikely in practice. Our method only requires \gls{los} to $4$ antennas, which in a large deployment with many antennas is likely even in the presence of some blockage. 
Methods that exploit multiple bands to resolve integer ambiguity have been proposed \cite{colocated_antennas,colocated_antennas_2,5g_phase_colocated, radio_interf_geolocation}, however, those rely on the assumption that the distance being measured at the different bands is identical. This is not possible at millimeter scale as the antennas cannot be physically placed in the same position. Contrary to that, our method enables accurate localization with arbitrary antenna deployments. 
True absolute localization with millimeter-level accuracy has only been achieved with specialized equipment, such as radars localizing modulated backscattering tags \cite{Hawkeye,backscatter_2} and sub-THz systems \cite{localization_100G}. These systems however require a significant additional complexity over a standard \gls{5g} deployment. 

\section{Discussion and future work} \label{sec:discussion}
In this section, we highlight several potential directions to further improve our system, and suggest methods and solutions to approach them. First, while coherent processing is becoming feasible with \gls{cran}, this  technology is not yet widely deployed. To enable carrier-phase localization without \gls{cran}, \gls{ota} phase synchronization needs to be explored and developed to sufficient accuracy. While this is well known to be challenging, we believe that the loose phase accuracy requirements of our method will be achievable with reasonable effort. Furthermore, the method requires sufficiently accurate baseband localization for initialization, which might not always be available. Thus, a natural improvement of the method would be to loosen the requirements on the initial estimate. To do so, in future work we plan to extend the method to explore multiple likelihood peaks at each step, and use a lightweight classifier to select the correct solution, based on the obtained likelihood values and contextual information such as previous estimates and other sensor inputs. Moreover, as quite accurate orientation information is required as part of the initial estimate, we will explore the use of multiple antennas at the TX side to obtain the initial orientation estimate, as well as to potentially refine it to $<0.1^\circ$ scale orientation accuracy (e.g., localizing two antennas spaced by $0.5$~m with a $500 \: \mu$m accuracy corresponds to a $0.057^\circ$ orientation error). Another avenue for improvement is simplifying the now labor-intensive calibration process. This can be done either by means of data-based approaches and machine learning, \edit{for example using a calibration wand with multiple \gls{tx} antennas at known distances, mimicking the procedure used for motion capture systems}, or by carefully modeling the sources of the impairments that are corrected by calibration. At present, the sources we have identified for such impairments are the imperfect knowledge of the antenna location and the variations of the antenna's phase center over different angles of arrival. Another seemingly straightforward extension of the work is to include millimeter-wave and sub-THz frequencies for micrometric accuracy. This, however, requires a precision around $100$ parts per million, where the limitation are no longer the radio or signal processing, but rather metrology issues such as air refractive index changes \cite{refractive_index_formula}, absolute clock accuracy and the crystal oscillator (e.g., some devices like \cite{clock_low_accuracy} exhibit frequency errors in this range) and thermal expansion of the building hosting the measurement system \cite{Wikipedia_CTE}. 
Finally, the method still requires $4$ (or $3$ in the case of $2$D localization) \gls{rx} antennas to work.
Using multipath reflectors as virtual anchors is a promising solution to reduce this number (or to further improve location accuracy). \edit{This is, however, very challenging because the reflection can be thought as the aggregate contribution of multiple infinitesimal surface currents across the reflector, and therefore is not happening in a well-defined small point (similarly to the antenna phase center issue). Thus, finding an equivalent reflection point and characterizing its response is only feasible if the reflector is known and present during calibration. Notably, this issue also applies to the case of device-free localization, where defining and characterizing the reflector at a sub-wavelength scale is a highly challenging problem.}

\section{Conclusions}
In this paper, we present the first millimetric accuracy \emph{one-shot absolute} positioning system for wireless communication networks, based on multi-band carrier phase measurement. The method has been extensively tested under challenging conditions, including the presence of strong multipath, antenna orientation changes, and the introduction of substantial synthetic noise. Under all conditions, the method achieves millimetric accuracy with high probability, outperforming classic localization methods by more than an order of magnitude. Furthermore, a novel \gls{fr3} hardware platform, used to evaluate the method in this paper, is made available to the community.

\bibliographystyle{acm}
\bibliography{references}

\end{document}